\documentclass[trackchanges, twocolumn]{aastex6}

\usepackage{braket}
\usepackage{subfigure}
\usepackage{graphicx}
\usepackage{amsmath}
\usepackage{amssymb}
\usepackage{physics}
\usepackage{tabularx}
\usepackage{mathtools}
\usepackage{mathdesign}

\makeatletter
\def\env@matrix{\hskip -\arraycolsep 
  \let\@ifnextchar\new@ifnextchar
  \array{*{\c@MaxMatrixCols}c}}
\makeatother

\begin{document}

\author{Tansu Daylan, Francis-Yan Cyr-Racine, Ana Diaz Rivero, Cora Dvorkin}
\affiliation{Department of Physics, Harvard University, Cambridge, MA, USA}
\email{tdaylan@fas.harvard.edu}
\author{Douglas P. Finkbeiner}
\affiliation{Department of Physics, Harvard University, Cambridge, MA, USA}
\affiliation{Harvard-Smithsonian Center for Astrophysics, Cambridge, MA, USA}

\received{16 June 2017}

\slugcomment{Submitted to \apj}

\title{Probing the small-scale structure in strongly lensed systems via transdimensional inference}

\begin{abstract}

Strong lensing is a sensitive probe of the small-scale density fluctuations in the Universe. We implement a pipeline to model strongly lensed systems using \emph{probabilistic cataloging}, which is a transdimensional, hierarchical, and Bayesian framework to sample from a \emph{metamodel} (union of models with different dimensionality) consistent with observed photon count maps. Probabilistic cataloging allows one to robustly characterize modeling covariances within and across lens models with different numbers of subhalos. Unlike traditional cataloging of subhalos, it does not require model subhalos to improve the goodness of fit above the detection threshold. Instead, it allows the exploitation of all information contained in the photon count maps -- for instance, when constraining the subhalo mass function. We further show that, by not including these small subhalos in the lens model, fixed-dimensional inference methods can significantly mismodel the data. Using a simulated Hubble Space Telescope (HST) data set, we show that the subhalo mass function can be probed even when many subhalos in the sample catalogs are individually below the detection threshold and would be absent in a traditional catalog. The implemented software, Probabilistic Cataloger (\texttt{PCAT}) is made publicly available at \url{https://github.com/tdaylan/pcat}.

\end{abstract}

\maketitle

\section{Introduction}
\label{sect:intr}

The concordance $\Lambda$ Cold Dark Matter ($\Lambda$CDM) model of cosmology is currently the most accurate description of matter density fluctuations in the Universe \citep{1981ApJ...250..423D,Blumenthal:1982mv,Blumenthal:1984bp,Davis:1985rj}. Through the collapse and bottom-up hierarchical merger of dark matter halos, it explains how structure could form in less than a billion years after the Big Bang. At the largest scales probed by the Cosmic Microwave Background (CMB), the six-parameter model agrees with precise observations of CMB anisotropies \citep{Ade:2015xua}. Nevertheless, discrepancies between observations and $\Lambda$CDM predictions have been claimed to remain at subgalactic scales \citep[see, e.g.][]{Bull2015, Popolo2016}. For instance, the number of observed satellites in the Local Group ($\sim$ 50) falls short of the much larger numbers of low-mass subhalos (> 1000) predicted by $N$-body simulations of $\Lambda$CDM \citep{Klypin1999,Moore:1999aa}. Furthermore, the projected density profiles of the observed subhalos are rather shallow in the central regions, compared to the much cuspier profiles predicted by $\Lambda$CDM \citep[see, e.g.][]{Flores:1994gz,Moore:1994yx,1995ApJ...447L..25B,2004MNRAS.351..903G,Walker:2011zu,2015AJ....149..180O}.

These apparent discrepancies can be resolved in several ways. First, the inclusion of baryonic feedback processes within numerical simulations \citep{2013ApJ...765...22B,2016MNRAS.457.1931S,2016ApJ...827L..23W,Despali2016,2017MNRAS.467.4383S,2017MNRAS.468.2283C,2017arXiv170103792G, Strigari2007, Nierenberg2016, Amorisco2014} has been shown to significantly reduce the tension between predictions and observations. Alternatively, the discrepancy could be due to a suppression of the small-scale matter power spectrum prior to structure formation. This may arise from the free streaming of dark matter (DM) particles after they kinetically decouple from the rest of the cosmic plasma, which partially erases inhomogeneities at the $\sim$ kpc comoving scale \citep{Bond:1983hb,Dalcanton:2000hn,Bode:2000gq,Boyanovsky:2008ab,Boyanovsky:2011aa}. Another venue to resolve the tension is to modify the microphysics of DM by, for instance, postulating that DM particles interact with each other \citep{Spergel2000,Yoshida00,Dave01,Colin02}, or with a relativistic species \citep[see, e.g.][]{1992ApJ...398...43C,Boehm:2001hm,Ackerman:2008gi,Feng:2009mn,Kaplan:2009de,Aarssen:2012fx,Cyr-Racine:2013ab,Chu:2014lja}. Both possibilities result in a modification to the internal density profile of DM halos, while the latter could also suppress the number of small-mass subhalos \citep[see, e.g.][]{Buckley:2014hja,Schewtschenko:2014fca,2016MNRAS.460.1399V}. Last, the problem could indicate a suppression of the \emph{primordial} power spectrum following inflation \citep{Kamionkowski:1999vp,Minor:2014xla}. In any case, the possibility of resolving the potential small-scale crisis of $\Lambda$CDM and exploring the particle nature of dark matter makes a strong case for observationally probing the subgalactic structure of DM.

Small-scale DM structure can, in principle, be probed by studying light emission from faint and ultra-faint dwarf galaxies in the Local Group and beyond. However, star formation in these low-mass subhalos is suppressed due to reionization, interaction with their host, or self-quenching \citep[e.g.][]{Fitts:2016usl,2017MNRAS.467.2019R}. Consequently most of the subhalos with virial halo mass $\lesssim 10^8M_\odot$ should be mostly dark, making their gravitational lensing \citep{Einstein1915} the only observable probe of their properties. Gravitational lensing thus provides a unique window into the smallest self-bound DM structure populating our Universe. In particular, galaxy-scale strong lensing systems in which a background source (e.g.~galaxy or quasar) is multiply imaged by a foreground massive galaxy provide a key way to probe the DM distribution on subgalactic scales.

Indeed, the observation of flux-ratio anomalies in strongly lensed quasars \citep{Mao:1998aa,Metcalf2001,Chiba:aa,Metcalf:ac, 2002astro.ph..2290D, Keeton:2003ab,Metcalf:2010aa, Fadely2011, Nierenberg:2014aa,gilman_strong_2017,2017arXiv170105188N, Nierenberg2017, 2017MNRAS.467.3970G, Hsueh2017} has been used to study the abundance of substructure within lens galaxies. In particular, \cite{Dalal2002} used the observation of seven radio-loud lensed quasars to put a constraint on the substructure mass fraction of $0.006 < f_{\rm sub} < 0.07$ (90\% confidence level). More recently, direct gravitational imaging of mass substructure \citep{Koopmans:aa,Vegetti:2008aa,Hezaveh:2012ai} lying close in projection to magnified arcs and Einstein rings has opened a new avenue to study the small-scale distribution of DM. This method works by fitting observed strong lenses with a smooth mass distribution, i.e. without substructure, and then adding pixel-based corrections \cite{Vegetti2010,Vegetti2012} or a number of linearized subhalos \cite{Hezaveh2016} to the mass model, with the requirement that the additional degrees of freedom improve the goodness of fit above some stringent detection threshold, e.g., greater than $\sim 10\sigma$. For instance, \cite{Vegetti2010,Vegetti2012} detected substructure with masses above $10^8 M_\odot$ at very high statistical significance using optical images of two strong lens systems. Similarly, a $10^9 M_\odot$ substructure was detected by \cite{Hezaveh2016} in an interferometric data set at a statistical significance of $6.9\sigma$ in the strong lens SDP.81 \citep{2015ApJ...808L...4A}. Constraints on the mass fraction in substructure and on the subhalo mass function can, in principle, be placed using these detections along with nondetections in other lens systems \citep[e.g.][]{2009MNRAS.400.1583V, Vegetti2014,Li:2015xpc,Hezaveh2016,Li:2016afu}. 

In practice, turning a direct mass substructure detection into constraints on the subhalo mass function and their internal properties (e.g.~concentration and truncation) is highly nontrivial. Indeed, degeneracies in the lensing model can blur the interpretation of a direct substructure detection \citep[e.g.][]{minor2016}. For instance, covariances between the smooth lens model, the structure of the source, and the subhalo population can lead to biases and underestimated error bars in the recovered substructure properties if they are not explored properly. Further, the nonlinearity of the lensing magnification implies that the likelihood improvement resulting from the addition of a subhalo to the lens mass model is not entirely independent of the presence of other subhalos within the lens. Although less likely, it remains a concern that two less massive subhalos serendipitously appearing in projection along the same line-of-sight could be misinterpreted as one more massive subhalo. Properly taking into account these degeneracies is important in order to obtain fully realistic error bars on subhalo population parameters (later identified as \emph{hyperparameters}). 

With the exception of several studies \citep[e.g.][]{2015ApJ...813..102B, Hezaveh2016a, Cyr-Racine2016, 2017JCAP...05..037B}, which aim to infer subhalos below the detection threshold \emph{statistically}, current analyses \citep[e.g.][]{Vegetti2014,Li:2015xpc,Hezaveh2016} place constraints on the subhalo population \emph{a posteriori} using highly statistically significant detections (or nondetections) of mass substructure within lens galaxies. In general, this requires writing down an approximate likelihood for the abundance of detected substructure, given some subhalo population parameters characterizing their mass function and spatial distribution. We identify two issues with this approach. First, by exclusively using an extremely reduced version of the data (i.e.~the position and mass of directly detected substructure), these analyses cannot take into account possible lensing degeneracies in which, for instance, the gravitationally imaged mass substructure actually corresponds to two close-by physical subhalos. Another way to phrase this problem is to say that correlations between models having different numbers of subhalos are neglected, which could potentially bias the recovered subhalo mass function constraints. Second, the use of highly reduced data neglects, by construction, marginal subhalo detections that, while not constituting statistically significant direct detections of substructure, can still provide statistical constraints on the subhalo population hyperparameters.

In this work, we develop a Bayesian data analysis framework that directly uses the lensed images as input and addresses the above issues by performing \emph{transdimensional} sampling over the union of lens models with different numbers of physical subhalos (hereafter, \emph{the metamodel}). Specifically, the lens metamodel refers to the union of mass models with a common macro (or smooth) lens model, and different numbers of subhalos, each with their own properties. The approach applies probabilistic cataloging \citep{Hogg2010, Brewer2012, Daylan2016, Brewer2015, Portillo2017} to images of strongly lensed systems. In contrast with standard gravitational imaging, our technique outputs an ensemble of probability-weighted subhalo catalogs providing good fits to the strongly lensed images. We use a version of the recently developed Markov chain Monte Carlo (MCMC) sampler, \software{Probabilistic Cataloger (\texttt{PCAT}) \citep{Daylan2016}}. \texttt{PCAT} is a Poisson mixture sampler that provides a transdimensional, hierarchical, and Bayesian inference framework to sample from the posterior probability distribution of a metamodel consistent with Poisson-distributed count data. In the context of the lensed images, we assume that the count data is collected by a photon-counting experiment such as a charged coupled device (CCD), but our results could, in principle, incorporate other types of auxiliary data (e.g., time delays of multiple images when working with quasar background sources \citep{Suyu2016}, or spectroscopic data of the foreground galaxy to constrain its mass budget \citep{Mason2017}), as well as informative priors on the relevant metamodel parameters.

Our approach shares its transdimensional nature with that of \citep{Brewer2015}, which implemented an independent transdimensional framework to infer substructure in strong lens systems. However, in addition to differences concerning substructure modeling and the sampling method used, our code and analysis were developed independently of theirs, and hence provide an important cross-check of the method and results. It has become customary in cosmological analyses to have at least two publicly available independent codes to check for numerical accuracy and find bugs. While both codes are public, we understand that this approach is not yet ubiquitous in the field of galaxy-scale strong lensing (for instance, it has been difficult thus far to reproduce/check the results of Vegetti et al. (2014) and Hezaveh et al. (2016)), but we hope that our work can help make substructure lensing more reproducible. Given the uniqueness of the constraints it can provide on low-mass substructures and the impact that these detections can have on dark matter physics, we believe it is extremely important to have multiple cross-checks to validate any results. On a more technical level, our approach differs from that of \citep{Brewer2015} in two major ways:
\begin{itemize}
\item Subhalo lens modeling. We represent subhalos with NFW profiles with a certain mass, scale, and cutoff radius, whereas \citep{Brewer2015} uses the so-called blobs, i.e., deflection as a power-law in the subhalo-centric radius. We think that choosing a physically motivated basis function set to represent the deflection field due to subhalos is essential for correctly marginalizing them out, i.e., propagating their (rather large) uncertainties to the parameter of interests such as the subhalo mass function.
\item The sampling method and the sampler used to sample from the posterior probability distribution of the (different) lens models given the data. \citep{Brewer2015} uses Diffusive Nested Sampling (DNS) as implemented in RJObject. We use a homebrew reversible-jump MCMC sampler with Metropolis-type within-model proposals.

\end{itemize}

Our work is also significantly different than that of \citep{Fadely2011}. In their analysis of the lens HE0435, they either consider models with a fixed number (1 to 3) of subhalos, thereby not taking into account the covariance between models with different numbers of subhalos, or they consider large populations of subhalos that are Poisson-distributed across the lens plane. Our approach of letting the data drive the actual number of subhalos is a significant improvement over this previous work.

After thoroughly sampling from the lens metamodel, one can marginalize over the subhalo catalogs to infer the underlying lens mass model and population characteristics of the subhalos, such as their mass function. The latter is achieved via hierarchical modeling, where the probability distributions of the subhalo properties, e.g., the subhalo mass function, are parameterized by \emph{hyperparameters}. When the subhalos in the metamodel are marginalized out, the posterior distribution of the hyperparameters constrains the population characteristics of the subhalos, even though most of the sampled subhalos are below the detection threshold. 

We base our inference directly on the observed photon count maps, as opposed to secondary, potentially biased estimates, such as astrometric perturbations or relative magnifications of the multiple images of the background source. This allows one to exploit all the information contained in the photon count maps regarding our metamodel, and to fully propagate uncertainties in the observed data to the parameter of interest, e.g., the subhalo mass function. 

To illustrate the features of probabilistic cataloging, we focus here on simple smooth lens models and source configurations, and present results for mock data sets; doing so allows one to study potential biases and systematics. Of course, analyses of real data would require more complex lens models -- and especially a more advanced lensing source reconstruction \citep[e.g.][]{Tagore:2014ssa}. While such complications are likely to make sampling from the metamodel more numerically onerous, they will not change the details of the probabilistic cataloging approach described below.

The rest of the paper is structured as follows. In Section \ref{sect:modl}, we discuss how we model lensed images. We then give an overview of probabilistic cataloging in Section \ref{sect:pcat}. We present our results using a mock data set in Section \ref{sect:resu} and assess the inference performance of probabilistic cataloging under mismodeling in Sections \ref{sect:null} and \ref{sect:minmdefs}. We discuss our results and conclude in Section \ref{sect:disc}.

\section{Lens metamodel}
\label{sect:modl}
In this section, we outline our lens metamodel and how we forward-model a strong lens image. The lensed image, $\tilde{f}_{\rm{src}}$, can be written as the source brightness profile $f_{\rm{src}}$, evaluated on the image plane,
\begin{align}
    \tilde{f}_{\rm{src}}(\theta_1, \theta_2) = f_{\rm{src}}(\theta_1 - \alpha_{\theta_1}, \theta_2 - \alpha_{\theta_2}),
    \label{equa:lensflux}
\end{align}
where $\vec{\alpha}=(\alpha_{\theta_1}, \alpha_{\theta_2})$ is the deflection vector at the image position, $\vec{\theta}=(\theta_1, \theta_2)$. Here, $\theta_1$ and $\theta_2$ denote the horizontal and vertical coordinates on the two-dimensional image plane, and $\vec{\tilde{\theta}}$ is the position on the source plane, such that
\begin{align}
    \vec{\tilde{\theta}} = \vec{\theta} - \vec{\alpha}(\vec{\theta}),
\end{align}
which is the lens equation. Furthermore, the deflection field $\vec{\alpha}$ can be written as the gradient of the Newtonian gravitational potential integrated along the line-of-sight,
\begin{align}
    \vec{\alpha} = \frac{2}{c^2} \vec{\nabla}_\theta \int \dd{\chi} \frac{\chi_{\rm{src}} - \chi}{\chi\chi_{\rm{src}}} \Psi(\vec{r}),
    \label{equa:deflcosm}
\end{align}
where $\chi$ is the comoving distance along the line-of-sight, $\chi_{\rm{src}}$ is the comoving distance to the source, and $\Psi(\vec{r})$ is the gravitational potential at the three-dimensional position vector, $\vec{r}$.

As in most lensing works, we make use of the thin lens approximation, whereby we assume that the thickness of the lensing halo is much smaller than other distance scales in the problem. This reduces Equation \eqref{equa:deflcosm} to

\begin{align}
    \vec{\alpha} = \frac{2}{c^2} \frac{D_{\rm{LS}}}{D_{\rm{S}}D_{\rm{L}}} \vec{\nabla}_\theta \int \dd{z} \Psi(\vec{r}),
    \label{equa:deflplan}
\end{align}
where $z$ is the line-of-sight coordinate, $D_{\rm{L}}$ and $D_{\rm{S}}$ are the angular diameter distances from the observer to the lens and source planes, respectively, and $D_{\rm{LS}}$ is the angular diameter distance from the lens to the source plane. Analytic expressions for the line-of-sight integral of various foreground mass components can then be used to calculate the deflection field \citep[see, e.g.][]{Keeton2001}.

\subsection{Deflection field}

Because of the linearity of the Poisson equation, the deflection field, $\vec{\alpha}$, can be expressed as the superposition of deflections due to several components. In a typical strong lens system, the dominant contribution comes from the smooth mass distribution of the main foreground galaxy (hereafter called the host halo), $\vec{\alpha}_{\rm{hst}}$ and the external shear due to the large-scale structure (LSS), $\vec{\alpha}_{\rm{ext}}$. Subhalos perturb this deflection field at the percent level, with individual contributions to the deflection field, $\vec{\alpha}_n$, where the subscript $n=1,2,..,N$ is the index of a subhalo. As a result, the total deflection field is given by
\begin{align}
    \vec{\alpha} = \vec{\alpha}_{\rm{ext}} + \vec{\alpha}_{\rm{hst}} + \sum_{n=1}^N \vec{\alpha}_n.
\end{align}
Let us now describe each component separately.

\paragraph{\textbf{Smooth halo}}

Cosmological $N$-body simulations of $\Lambda$CDM \citep{Genel2014} suggest that the equilibrium distribution of collisionless dark matter particles is approximately spherically symmetric about the dynamical center of mass of the self-gravitating halo of dark matter particles. Furthermore, the resulting virialized halos have a universal radial profile that is well-described by a broken power-law in the radial halo-centric distance, $r$, and is known as the Navarro Frenk White (NFW) profile \citep{Dubinski1991, Navarro1996}. However, the host halo also possesses baryonic matter, which actually dominates the mass budget in the bulge. Taking the mass budget of the host halo to be the sum of the baryonic and DM components, it has been found that the overall mass density profile of early-type galaxies (ETGs) can be fitted well by singular isothermal ellipsoids (SIEs), which is known as the bulge-halo conspiracy \citep{Treu2010} because neither dark nor baryonic matter density distribution is individually isothermal.

We therefore take the mass density profile of the host halo as
\begin{align}
    \rho_{\rm{hst}}(r) = \frac{\sigma^2_{\rm{hst}}}{2\pi G r^2},
\end{align}
where $\sigma_{\rm{hst}}$ is the (constant) velocity dispersion and $G$ is the gravitational constant. In general, galaxies also possess some ellipticity due to the existence of a baryonic disc or as artifacts of recent gravitational interactions. By taking into account the ellipticity of the isothermal gas, and assuming that the contours of equal surface mass density are in the form of concentric ellipses, the deflection is given by \citep{Keeton2001, Kassiola1993},
\begin{multline}
    \vec{\alpha}_{\rm{hst}} = \frac{\theta_{\rm{E,hst}} q_{\rm{hst}}}{\sqrt{1 - q_{\rm{hst}}^2}} \Bigg( 
    \tan^{-1}\Big(\frac{\sqrt{1 - q_{\rm{hst}}^2} \theta_{1,\rm{hst}}^\prime}{\omega} \Big) \hat{\theta}_{1,\rm{hst}}^\prime + \\
    \tanh^{-1}\Big(\frac{\sqrt{1 - q_{\rm{hst}}^2} \theta_{2,\rm{hst}}^\prime}{\omega} \Big) \hat{\theta}_{2,\rm{hst}}^\prime
    \Bigg),
    \label{equa:hostdefl}
\end{multline}
where $\omega\equiv \sqrt{q_{\rm{hst}}^2 {\theta_1^\prime}^2 + {\theta_2^\prime}^2}$, $q_{\rm{hst}} \equiv 1 - \epsilon_{\rm{hst}}$ is the minor to major axis ratio, and $\epsilon_{\rm{hst}}$ is the ellipticity. Note that the primed quantities, i.e, $\theta_{1,\rm{hst}}^\prime$, $\theta_{2,\rm{hst}}^\prime$, $\hat{\theta}_{1,\rm{hst}}^\prime$, and $\hat{\theta}_{2,\rm{hst}}^\prime$, denote the angular coordinates and the associated unit vectors along the minor and major axes of the elliptical host halo, respectively. If we use $\phi_{\rm hst}$ to denote the angle between the major axis of the ellipse and the horizontal axis of the image, $\theta_1$, then the primed coordinates are obtained by translating the image by $(\theta_{1,\rm{hst}}, \theta_{2,\rm{hst}}$) and then rotating it by $\phi_{\rm{hst}}$. 

Furthermore, the lensing strength of the isothermal deflector is parameterized using the projected Einstein radius, $\theta_{E,\rm{hst}}$, i.e., the radius of the circular image that would be produced if the lens were spherical and exactly aligned with the background light source. In a more general sense that holds for non-spherical lenses, we define $\theta_{E,\rm{hst}}$ as the radius of a circle that has an area equal to the area inside the critical curve.

Note that many observed strong lenses require departures from an isothermal model, where the preferred inner log-slope of the three-dimensional density distribution can be significantly different from -2. However, because the method we present is independent of the details of the lens model, we assume a relatively simple mass model for the smooth halo, leaving further refinements to future work.

\paragraph{\textbf{Subhalos}}

$\Lambda$CDM predicts that, with cosmological time, more density peaks collapse to form new self-gravitating halos, while halos that formed earlier are accreted into more massive ones. Because hierarchical structure formation is a continuous and ongoing process, at any given time some fraction of matter is expected to be tied in halos that have been accreted into more massive halos. In the rest of the paper, we refer to these dark, gravitationally bound satellites inside the virial extent of the host as \emph{subhalos}.

Subhalos grow by accretion and lose mass via tidal stripping, collisions, and evaporation. The relative efficiency of these processes determines the level of substructure in a host halo of mass $M_{\rm{hst}}$ and redshift $z_{\rm{hst}}$. $N$-body simulations can predict the abundance and properties of these subhalos, although the finite mass resolution of simulations can preclude robust conclusions \citep{Bosch2016}. Nevertheless, it is expected that subhalos have a truncated mass profile due to tidal stripping by the host galaxy. Therefore, we take the three-dimensional subhalo density profile of the $n$th subhalo, $\rho_n(r)$, to be the NFW profile with a power-law truncation as \citep{Baltz2009}:
\begin{align}
\rho_n(r) = \frac{M_{0,n}}{4\pi r_{\rm{s}, \textit{n}}^3} \frac{1}{(r/r_{\rm{s}, \textit{n}})(1 + r/r_{\rm{s}, \textit{n}})^2} \frac{\tau_n^2}{(r/r_{\rm{s}, \textit{n}})^2+\tau_n^2}, \\
M_{\rm{c},\textit{n}} = M_{0,n} \frac{\tau_n^2}{(\tau_n^2 + 1)^2} \left[ (\tau_n^2 - 1) \ln \tau_n + \pi \tau_n - (1 + \tau_n^2)\right].
\end{align}
where $M_{\rm{c},\textit{n}}$ is the truncated mass of the same subhalo. In this parameterization most of the mass of a subhalo is contained inside the scale radius, $r_{\rm{s},\textit{n}}$, which would be the subhalo-centric distance at which the density profile steepens from a power-law with index -1 to a power-law with index -3 if it was not tidally truncated. Moreover, the central mass concentration is encoded in $\tau_n=r_{\rm{c},\textit{n}}/r_{\rm{s},\textit{n}}$ similarly to the concentration parameter $c=r_{\rm{vir},\textit{n}}/r_{\rm{s},\textit{n}}$ in the non-truncated case, where $r_{\rm{vir},\textit{n}}$, $r_{\rm{c},\textit{n}}$ and $r_{\rm{s},\textit{n}}$ are the virial, tidal cutoff, and scale radii of a subhalo. $M_{0,n}$ is again a mass scale relevant to the $n$th subhalo. Subhalos with large $\tau_{n}$ represent rather old, highly concentrated subhalos that orbit the host at large halocentric distances. In contrast, those with low $\tau_n$ typically correspond to subhalos that are close to the center and are thus being tidally stripped by the host. Note that we use the subscript $n$ when referring to a parameter of the $n$th subhalo when the host halo has the same type of parameter.

It is also expected that subhalos near the dynamical center of the host halo, and thus closest to disruption, should be tidally elongated. Nevertheless, because current data sets do not have the signal-to-noise ratio (SNR) necessary to constrain departures from spherical symmetry, we neglect any triaxiality in the subhalo shapes and assume that they are spherically symmetric. Furthermore, because virialized halos are generally found to have concentrations that reflect the average background density of the Universe at the time of their collapse, earlier halos are expected to have higher concentrations. Upon accretion into the host halo, their high concentrations can help them remain intact despite strong tidal stretching and evaporation by the host halo. However, motivated by recent $N$-body simulations \citep{Springel2008}, we neglect substructure inside subhalos and assume that the subhalos of the host galaxy have smooth mass distributions.

Although the mass of a non-truncated NFW profile is logarithmically divergent with halocentric radius, tidal truncation makes the total mass $M_{\rm{c},\textit{n}}$ finite \citep{Baltz2009} such that
\begin{align}
M_{\rm{c},\textit{n}} = M_{0,n} \frac{\tau_n^2}{(\tau_n^2 + 1)^2} \left[ (\tau_n^2 - 1) \ln \tau_n + \pi \tau_n - (1 + \tau_n^2)\right].
\end{align}

\begin{figure}
    \includegraphics[width=0.45\textwidth]{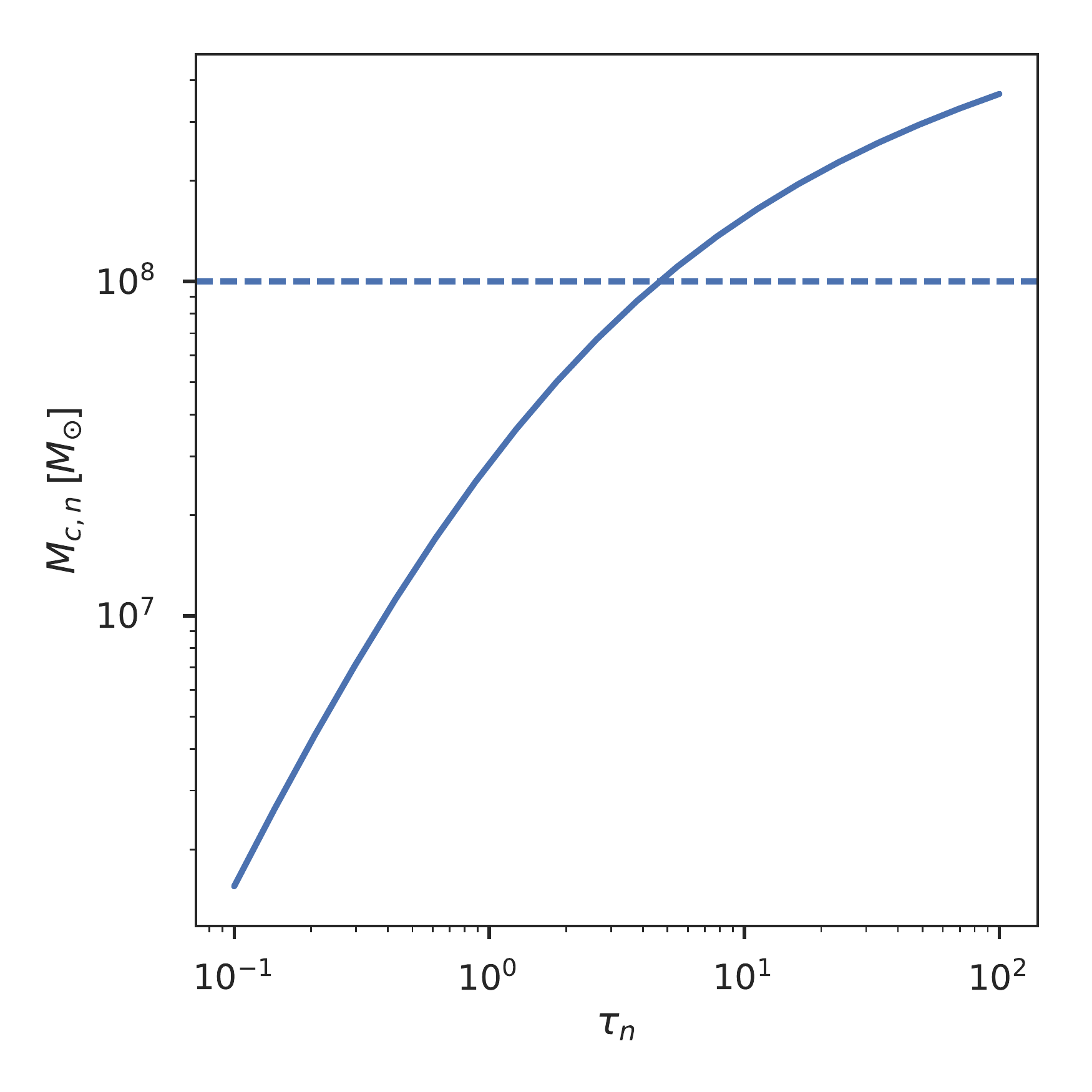}
    \caption{Truncated mass of a subhalo, $M_{\rm{c},\textit{n}}$, normalized by $M_{0,n}$ as a function of the ratio between its cutoff and scale radii, $\tau_n$.}
    \label{figr:mcut}
\end{figure}

This finite mass is plotted, in Figure \ref{figr:mcut} for a subhalo in units of $M_{0,n}$. It changes by about an order of magnitude in the relevant region of $1\lesssim \tau_n \lesssim 10$. Below $\tau_n \sim 1$, the subhalo is likely to be completely disrupted by evaporation and tidal stripping. We report truncated masses when referring to subhalos. 

Our simulated images are approximately centered at the dynamical center of the host halo and have a size comparable to its Einstein radius, which is much smaller than its virial radius. Because we work on the (projected) lens plane, subhalos in the field of view have a large spread in their three-dimensional distance from the dynamical center of the host. As a result, there should be a large spread in their deflection profile. We crudely take this into account by drawing the projected scale and cutoff radii of the mock subhalos from a uniform distribution. The limits of these distributions are given in Table \ref{tabl:modl}. The resulting ratios between the projected cutoff and scale radii, $\tau_n$, fall between 3 and 25.

We emphasize that these mock subhalos are meant to provide a test bed for introducing probabilistic cataloging of subhalos and do not fully reflect their rich dynamics. Furthermore, although simulations and analytical treatments of $\Lambda$CDM can predict the scale and cutoff radii of a given subhalo, if provided its mass and distance from the center of the host galaxy \citep{Metcalf2001}, which implies that predictions can be folded into the inference as a joint prior distribution on the projected scale and cutoff radii, the fact that strong lens images are not informative about the three-dimensional distance of a subhalo from the center of its host still means there is a degeneracy in this description.

We parameterize the deflection field due to the $n$th subhalo, $\alpha_n$, using the angular distances corresponding to the scale and cutoff radii, $\theta_{\rm{s},\textit{n}}=r_{\rm{s},\textit{n}}/D_{\rm{L}}$ and $\theta_{\rm{c},\textit{n}}=r_{\rm{c},\textit{n}}/D_{\rm{L}}$, and the normalization of the deflection profile, $\alpha_{\rm{s},\textit{n}}$, which we refer to as the deflection strength. The functional form of this deflection profile is presented in Appendix \ref{sect:deflsubh}. 

\begin{figure}
    \includegraphics[width=0.45\textwidth]{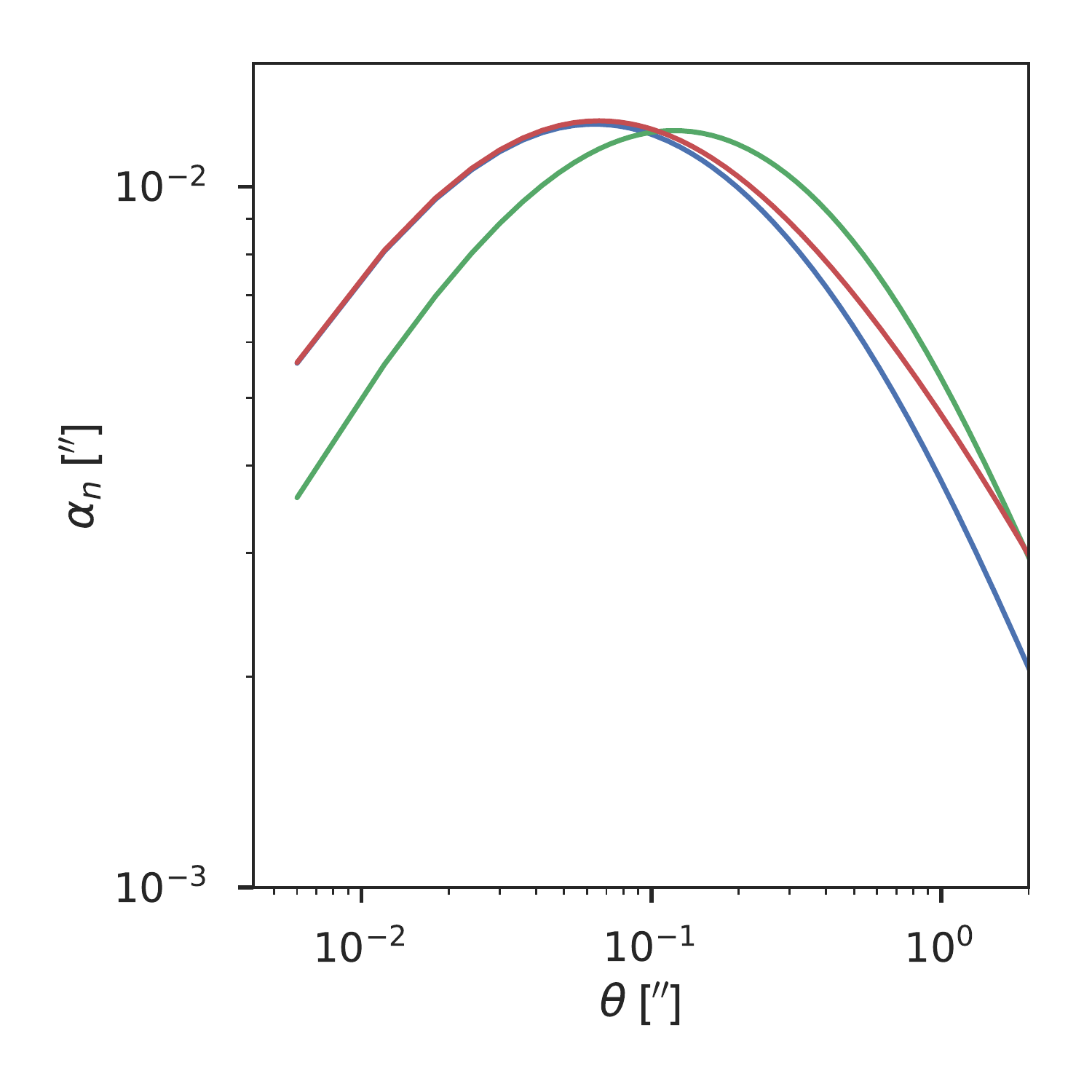}
    \caption{Deflection as a function of angular distance from a given subhalo in arcsec. All three profiles have a subhalo deflection strength of $\alpha_{\rm{s},\textit{n}} = 0.1$ arcsec. The blue, green, and red profiles correspond to subhalos with ($\theta_{\rm{s},\textit{n}}= 0.05$ arcsec, $\theta_{\rm{c},\textit{n}} = 1$ arcsec), ($\theta_{{\rm s},n}= 0.2$ arcsec, $\theta_{{\rm c},n} = 1$ arcsec) and ($\theta_{\rm{s},\textit{n}}= 0.05$ arcsec, $\theta_{\rm{c},\textit{n}} = 2$ arcsec), respectively.}
    \label{figr:deflcutf}
\end{figure}

The deflection as a function of radial distance in three dimensions from the dynamical center of mass of the subhalo is proportional to the integrated mass up to that radius. Therefore, it initially rises because of the shallow log-slope of the NFW profile, turns over at the scale radius, and further steepens beyond the cutoff radius. The projected deflection profile of a subhalo is given in Figure \ref{figr:deflcutf}.

$\Lambda$CDM predicts that the variance of the matter density fluctuations in the Universe monotonically increases as one averages the density field inside successively smaller spheres. Therefore, the number of collapsed halos (and hence of subhalos) increases steeply as a power-law toward low masses with a log-slope of $\sim -1.9$ \citep{Springel2008}. When $\Lambda$CDM is modified at small scales via self-interactions, free-streaming at kinetic decoupling, or by the Heisenberg uncertainty in the case of wave-like dark matter \citep{Hu2000}, this is no longer the case and one expects a mass scale below which the subhalo mass function is suppressed.

Because subhalo deflection strengths are proportional to mass, we assume $\alpha_{\rm{s},\textit{n}}$ are also distributed as a power-law between some minimum and maximum value. However, note that the lower limit of this prior does not necessarily correspond to a physical cutoff in the mass distribution. Rather, it is a computational convenience, because probabilistic cataloging of very low-mass subhalos results in very little information gain -- at the expense of significantly increased computational cost. Furthermore, the pixel size and the point-spread function (PSF) of the photon-collecting device sets an angular distance scale below which deflections cannot significantly affect the image. This scale also depends on the level of background emission and the flux of the lensed light source. For our simulated images of the Wide Field Camera 3 / Ultraviolet Visible (WFC3/UVIS) camera on board the Hubble Space Telescope (HST), whose pixel size is 0.04 arcsec, this scale roughly corresponds to $\sim$ 0.01 arcsec. We therefore choose the minimum allowed value of $\alpha_{\rm{s},\textit{n}}$ as 0.01 arcsec, in order to allow low-significance subhalos into the lens metamodel. We also choose the maximum of $\alpha_{\rm{s},\textit{n}}$ to be 1 arcsec. However, this maximum value does not affect our prior on $\alpha_{\rm{s},\textit{n}}$ because most of the prior volume is contained at small $\alpha_{\rm{s},\textit{n}}$ due to the steepness of the power-law. The normalization of this power-law is then given by $\mu_{\rm{sub}}$.

In the hierarchical modeling approach of probabilistic cataloging, the negative of the log-slope of the $\alpha_{\rm{s},\textit{n}}$ distribution, $\beta$, is also allowed to vary. In this inference scheme, the sampler visits all \emph{prior} configurations consistent with the data and generates a posterior distribution over $\beta$. Therefore, although $\beta$ parameterizes the prior, it is treated just like any other parameter in the model, and is referred to as a \emph{hyperparameter}. It is the prior on this \emph{hyperparameter} that specifies our prior belief on the distribution of $\alpha_{\rm{s},\textit{n}}$. We set this prior such that $\beta$ itself is Gaussian-distributed with a mean and standard deviation of 1.9 and 0.5, respectively. The mean reflects the value motivated by $\Lambda$CDM, whereas the spread accounts for any deviations and allows the sampler to visit subhalo configurations that deviate from a power-law with index 1.9.

\begin{figure}
    \includegraphics[width=0.45\textwidth]{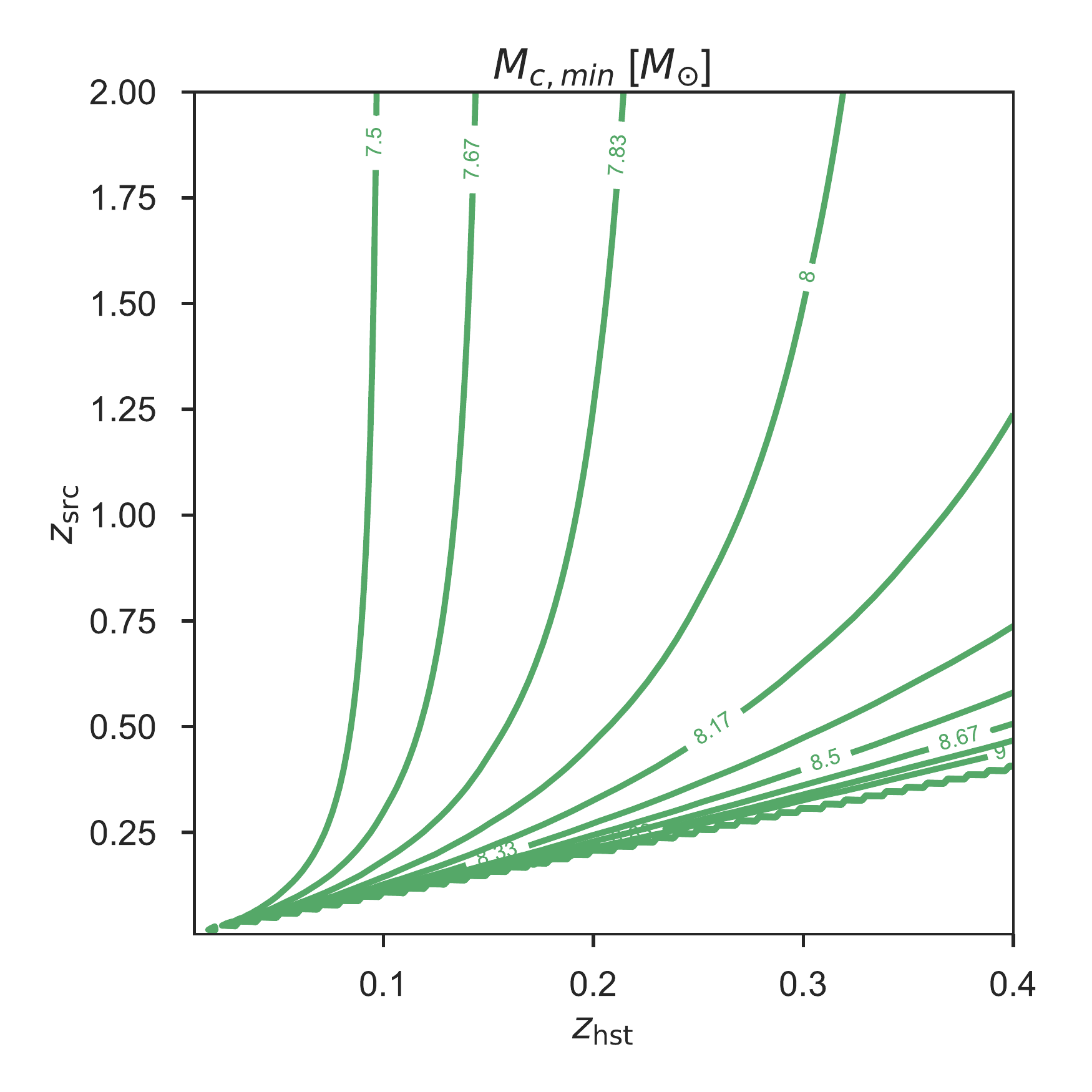}
    \caption{Contours of minimum truncated subhalo mass allowed in the metamodel, i.e., lower limit of the truncated subhalo mass prior, as a function of host and source galaxy redshifts, assuming $\theta_{\rm{s},n}=0.1$ arcsec and $\theta_{\rm{c},n}=1$ arcsec.}
    \label{figr:massredsminm}
\end{figure}

The deflection strength, $\alpha_{\rm{s},\textit{n}}$, and projected scale and cutoff radii, $\theta_{\rm{s},\textit{n}}$ and $\theta_{\rm{c},\textit{n}}$, of a subhalo uniquely determine its truncated mass, $M_{\rm{c},\textit{n}}$. Therefore, given the above priors, there is no hard boundary on the prior distribution of the subhalo truncated masses. However, most of the prior volume falls in the interval $10^8 - 10^9 M_\odot$ for the fiducial host and source redshifts of $z_{\rm{hst}} = 0.2$ and $z_{\rm{src}} = 1$. This interval changes as a function of angular diameter distances to the host and source planes. To put our prior mass interval into cosmological context, Figure \ref{figr:massredsminm} shows the lower bound of the prior on the truncated mass of a subhalo, if it had $\theta_{\rm{s},\textit{n}}= 0.05$ arcsec and $\theta_{\rm{c},\textit{n}} = 1$ arcsec.

Finally, we adopt a spatially uniform prior on the projected surface density of subhalos. The prior structure of our lens metamodel is summarized in Table \ref{tabl:modl}.

\paragraph{\textbf{External shear}}

Lensing galaxies are usually found in galaxy groups or clusters whose overall contribution adds angular structure to the deflection field in the vicinity of the lensing galaxy. Even in the case of an isolated galaxy, matter distribution in the foreground of the lensing galaxy can introduce additional multipole terms to the deflection field. We parameterize this external deflection by means of a reduced shear field $\gamma_{\rm{ext}}$ oriented at an angle $\phi_{\rm{ext}}$ with respect to the longitudinal axis \citep{Keeton1997}. The field is traceless and invariant under a rotation of 180$^\circ$.

\begin{align}
    \vec{\alpha}_{\rm{ext}} = \gamma_{\rm{ext}}
    \begin{pmatrix}
        \cos 2\phi_{\rm{ext}} &  + \sin 2\phi_{\rm{ext}} \\
        \sin 2\phi_{\rm{ext}} &  - \cos 2\phi_{\rm{ext}}
    \end{pmatrix} \vec{\theta}.
\end{align}
Note that we only model the reduced shear, $\gamma_{\rm{ext}}=\gamma_{\rm{ext}}^\prime/(1 - \kappa_{\rm{ext}})$, where $\kappa_{\rm{ext}}$ represents a potentially underlying, spatially uniform convergence in the field of interest. This is because adding a constant mass sheet to the model, $(1 - \kappa_{\rm{ext}})$, reproduces the same lensed image by simultaneously rescaling the source position and flux or the deflector mass \citep{Falco1985}. These degenerate mass models are linked by the transformations
\begin{align}
\begin{split}
    (1 - \kappa_{\rm{ext}}) &\to \lambda \kappa_{\rm{ext}}, \\
    \gamma_{\rm{ext}} &\to \lambda \gamma_{\rm{ext}}.
    \label{equa:massdege}
\end{split}
\end{align}
An independent measurement of the deflector mass, e.g., using stellar kinematics, could constrain $\lambda$ \citep{Schneider2013, Bradac2004}. Alternatively, because the magnification field and light travel-time change with $\lambda$, knowledge of the non-lensed fluxes of the background sources or having multiple images of the same system taken at different times could break this degeneracy. Specifically, the knowledge of time delays between multiple images of a background quasar and the time-delay distance allows one to constrain the gravitational potential of the main deflector, independent of the astrometric information that we use in this work. Because we only rely on a single photometric exposure of the strong lens system, we simply determine all convergences up to the transformation in Equation \ref{equa:massdege} and do not attempt to lift the degeneracy.

We note that the inclusion of only a quadrupolar shear field is motivated by the assumption that the lensing galaxy is isolated and gravitationally relaxed. If there are nearby galaxies, higher-order multipoles would be needed.

\subsection{Light emission from the source}

In general, the background light source in a strong lens system could be: a quasar; a blue, young galaxy with bright Lyman-$\alpha$ emission; or a red ETG similar to the lensing galaxy we consider in this work. When the background light source is a quasar, i.e., a point source without any spatial feature, the resulting multiple images are not extended and can be used to probe subhalos only over a small region on the lens plane depending on the mass of the subhalo. We therefore instead consider galaxy-galaxy lenses, where light from a background galaxy is strongly deflected by a giant elliptical in the foreground. In these systems, multiple images are more extended and a higher fraction of the pixels are informative on the subhalo parameters.

We assume that the emission from both the foreground and the background galaxies have S\'ersic profiles, i.e., the surface brightness is given by \citep{1963BAAA....6...41S}
\begin{align}
    f_{\rm{gal}}(\theta) = f_{\rm{e,gal}} \exp\Bigg( -b_n\Big((\theta/\theta_{\rm{e,gal}})^{1/n} - 1\Big) \Bigg),
    \label{equa:sersprof}
\end{align}
where the subscript gal refers to both lensing (foreground) and lensed (background) galaxies, i.e., gal=\{hst,src\}. In this relation, $b_n$ is a coefficient that depends on the index $n$ \citep{Ciotti1999},
\begin{align}
    b_n = 2n - \frac{1}{3} + \frac{4}{405n} + \frac{46}{25515n^2}
\end{align}
which controls the level at which the inner and outer slopes anti-correlate. Furthermore, $\theta_{\rm{e,gal}}$ is the projected distance within which half of the total emission is contained and $f_{\rm{e,gal}}$ is the surface brightness at this radius. We assume $n=4$, which corresponds to the de Vaucouleurs profile \citep{Vaucouleurs1948}. For this profile, the surface brightness at the half-light radius, $f_{\rm{e,gal}}$, is related to the flux of the galaxy, $I_{\rm{gal}}$, by $I_{\rm{gal}} \approx 7.2 \pi \theta_{\rm{e},gal}^2 f_{\rm{e,gal}}$.

Modeling the source plane emission with the surface brightness of a single galaxy can be a large simplification compared to real lenses. In particular, emission or absorption regions in the source galaxy can significantly change the appearance of the resulting arcs and arclets. However, in this work we aim to construct a simple test bed to compare probabilistic cataloging to mainstream analysis tools of strong lensing. Therefore, we leave a pixel-based representation of the source plane to future work.

\subsection{Light emission from the foreground galaxy}

We use the same surface brightness profile in Equation \ref{equa:sersprof} to represent emission from the host galaxy, but with a different half-light radius and surface brightness, $\theta_{\rm{e,hst}}$, and $f_{\rm{e,hst}}$.

Because the scattering cross section between dark and baryonic matter is not necessarily zero, in principle there can be a spatial offset between the surface brightness and mass density of a host galaxy that has not been gravitationally relaxed \citep{Shu2016}. We do not model such unrelaxed systems, and assume that the smooth mass distribution and the light emission of the host galaxy are co-centric.

Furthermore, we do not subtract an estimate of foreground galaxy or isotropic emission from the observed photon-count map. This is because subtracting a best-fit emission component from the photon count map is equivalent to placing a delta function prior at the associated value, or equivalently, assuming that the parameter has vanishing covariance with all other parameters. Instead, we sample from the joint posterior probability distribution of all parameters including those of the host galaxy, and later marginalize over the nuisance parameters.

\subsection{Background emission}

Any CCD image includes a spatially uniform contribution from instrumental and astronomical backgrounds. We model this by adding an isotropic emission template with a floating normalization to the predicted images. This template also absorbs any true sky background in the vicinity of the strong lensed system. Note that the shot noise is built into the model by evaluating the Poisson probability of realizing the number of counts in the observed image, given the number of photons predicted by the model.

\subsection{Instrumental PSF}

Due to atmospheric distortions or imperfections in the optics, an optical photon-collecting device spreads an input delta function to a characteristic shape on the image plane, known as the PSF. In general, this shape can be modeled as a linear combination of Gaussians or (when the PSF has heavy tails) Lorentzians. A frequently used distribution is that of Student's t-distribution (also known as the Moffat profile), which can account for both the Gaussian core and the heavy tails of the PSF shape.

In this paper, we simulate an HST image. In the absence of an intervening atmosphere and within the limit of distortion-free optics, we assume that the PSF is instead limited by diffraction, although the real HST PSF has diffraction spikes due to the support vanes of the secondary mirror. Given the circular aperture of the HST, we model the PSF in band $i$ as an Airy pattern whose first zero-crossing occurs at a radius of $\sigma_{\textrm{psf},i}$. 

In general, uninformative priors can be placed on the parameters that characterize the PSF shape, e.g., $\sigma_{\textrm{psf},i}$. This would allow PSF parameters to be inferred along with the lens metamodel. However, in the absence of a bright point source in the image, the PSF cannot be constrained-well. In fact, for a given telescope, the shape of the PSF is usually well-constrained using independent calibration data. Therefore, \texttt{PCAT} assumes Gaussian priors on the PSF parameters, whose means and standard deviations are provided by the external calibration. This allows exploitation of previous calibration while also still allowing propagation of uncertainties due to potential covariances between the PSF and model parameters.

\subsection{Model image}

Given the above ingredients, we express the model image (prior to the PSF convolution) as the sum of
\begin{enumerate}
\item a spatially uniform template to model detector background and isotropic sky emission, $f_{\rm{bac}}$,
\item emission from the host galaxy, $f_{\rm{hst}}$,
\item the gravitationally lensed emission of a background light source due to a foreground host galaxy and a variable number of its subhalos, $\tilde{f}_{\rm{src}}$.
\end{enumerate}
The total model emission, $f_{\textrm{m}}$, is then obtained by convolving this image with the PSF kernel, $\mathcal{P}$, such that
\begin{align}
    f_{\textrm{m}} = \mathcal{P} * (f_{\rm{bac}} + f_{\rm{hst}} + \tilde{f}_{\rm{src}}).
\end{align}

\subsection{Model indicator}

Our lens metamodel is a union of lens models, each with a fixed number of subhalos. This implies that the number of subhalos, $N_{\rm{sub}}$, which is also known as the model indicator, is a discrete parameter of the metamodel. Through transdimensional proposals that respect detailed balance, it gets updated as the Markov chain state evolves, generating a posterior distribution over $N_{\rm{sub}}$. We assume that $N_{\rm{sub}}$ is Poisson-distributed with a mean of $\mu_{\rm{sub}}$ and place a uniform prior on the distribution of $\mu_{\rm{sub}}$. Therefore, in principle, $N_{\rm{sub}}$ can be any non-negative integer. However, we place an upper limit of 100 on it, for computational convenience, and check that this upper limit is never reached. Implementation of transdimensional proposals in \texttt{PCAT} is discussed in Section \ref{sect:pcat}.

\subsection{Summary of the lens metamodel}
\label{sect:modlsumm}

Table \ref{tabl:modl} lists the metamodel parameters. The group of rows at the top contains the hyperparameters, which parameterize the conditional distribution of subhalo properties. These subhalo parameters are shown in the group of rows at the bottom of the list. The remaining groups of rows are the PSF, background, and lens parameters, from top to bottom. The lens parameters are further divided into three subgroups, separately showing the foreground host lens, background light source, and external shear parameters.

The third and fourth columns indicate the minimum and maximum when the associated distribution is uniform, log-uniform or power-law, and show the mean and standard deviation when the distribution is Gaussian. The fifth column indicates the parameters of the metamodel from which our mock data sets are drawn. Generation of mock data sets will be discussed in Section \ref{sect:mock}. 

\begin{table*}
\caption{Parameter List of the \texttt{PCAT} Lens Metamodel}
\centering
\begin{tabular}{|c||c|c|c|c|c|c|c}
    \hline
    name                     & prior          & min/mean            & max/std                 & true            & unit                & explanation \\ 
    \hline
    \hline
    \hline
    \hline
    $\mu_{\rm{sub}}$         & uniform        & 0                   & 100                     & -               &                     & mean number of subhalos \\
	$\beta$                  & Gaussian       & 1.9                 & 0.5                     & 1.9             &                     & slope of the deflection strength distribution \\
    \hline
    \hline
	$\sigma_{\textrm{psf}}$  & Gaussian       & 0.087               & 0.01                    & 0.087           & arcsec              & radius of the Airy disk that represents the PSF \\
    \hline
    \hline
	$A$                      & $\log$-uniform & $10^{-8}$           & $10^{-6}$               & $2\times10^{-7}$& erg/cm$^2$/s/\AA/sr & normalization of the isotropic background emission \\
    \hline
    \hline
	$\theta_{1,\rm{src}}$    & uniform        & -2                  & 2                       & RC              & arcsec              & horizontal coordinate of the background source \\
	$\theta_{2,\rm{src}}$    & uniform        & -2                  & 2                       & RC              & arcsec              & vertical coordinate of the background source\\
	$I_{\rm{src}}$           & $\log$-uniform & $10^{-20}$          & $10^{-15}$              &$10^{-18}$       & erg/cm$^2$/s/\AA    & flux of the background source \\
	$\theta_{e,\rm{src}}$    & $\log$-uniform & 0.1                 & 2                       & 0.5             & arcsec              & half-light radius of the background source \\
	$\epsilon_{\rm{src}}$    & uniform        & 0                   & 0.3                     & R               &                     & ellipticity of the background source \\
	$\phi_{\rm{src}}$        & uniform        & 0                   & 2$\pi$                  & R               & radian              & azimuthal orientation of the background source \\
    \hline
	$\theta_{1,\rm{hst}}$    & uniform        & -2                  & 2                       & RC              & arcsec              & horizontal coordinate of the host galaxy \\
	$\theta_{2,\rm{hst}}$    & uniform        & -2                  & 2                       & RC              & arcsec              & vertical coordinate of the host galaxy \\
	$I_{\rm{hst}}$           & $\log$-uniform & $10^{-20}$          & $10^{-15}$              &$10^{-16}$       & erg/cm$^2$/s/\AA    & flux of the host galaxy \\
	$\theta_{\rm{e,hst}}$    & $\log$-uniform & 0.1                 & 2                       & 1               & arcsec              & scale size of the S\'ersic profile of the host galaxy \\
	$\theta_{\rm{E,hst}}$    & $\log$-uniform & 0.5                 & 2                       & 1.5             & arcsec              & Einstein radius of the host galaxy \\
	$\epsilon_{\rm{hst}}$    & uniform        & 0                   & 0.5                     & R               &                     & ellipticity of the host galaxy \\
	$\phi_{\rm{hst}}$        & uniform        & 0                   & 2$\pi$                  & R               & radian              & azimuthal angle of the ellipticity of the host galaxy \\
    \hline
	$\gamma_{\rm{ext}}$      & uniform        & 0                   & 0.3                     & R               &                     & amplitude of the external shear \\
	$\phi_{\rm{ext}}$        & uniform        & 0                   & 2$\pi$                  & R               & radian              & azimuthal angle of the external shear \\
    \hline
    \hline
    $N_{\rm{sub}}$           & Poisson        & $\mu_{\rm{sub}}$    & $\sqrt{\mu_{\rm{sub}}}$ & 25              &                     & number of subhalos in a model \\
	$\vec{\theta_{1}}$       & uniform        & -2                  & 2                       & R               & arcsec              & horizontal coordinate of the $n$th subhalo \\
	$\vec{\theta_{2}}$       & uniform        & -2                  & 2                       & R               & arcsec              & vertical coordinate of the $n$th subhalo \\
	$\vec{\alpha_{\rm{s}}}$  & power-law      & 0.01                & 1                       & R               & arcsec              & deflection strength of the $n$th subhalo \\
	$\vec{\theta_{\rm{s}}}$  & uniform        & 0                   & 0.1                     & R               & arcsec              & projected scale radius of the $n$th subhalo \\
	$\vec{\theta_{\rm{c}}}$  & uniform        & 0                   & 2                       & R               & arcsec              & projected cutoff radius of the $n$th subhalo \\
    \hline
\end{tabular}
\textbf{Note.} The letter R under the ``true'' column implies that the associated parameter is randomly sampled from the prior. The letters RC, on the other hand, mean that the parameter was drawn from a Gaussian with mean 0 and standard deviation 0.04 arcsec. See the text for other details.
\label{tabl:modl}
\end{table*}

\begin{figure*}
    \centering
    \includegraphics[width=0.55\textwidth, trim=3cm 2cm 1cm 1cm, clip]{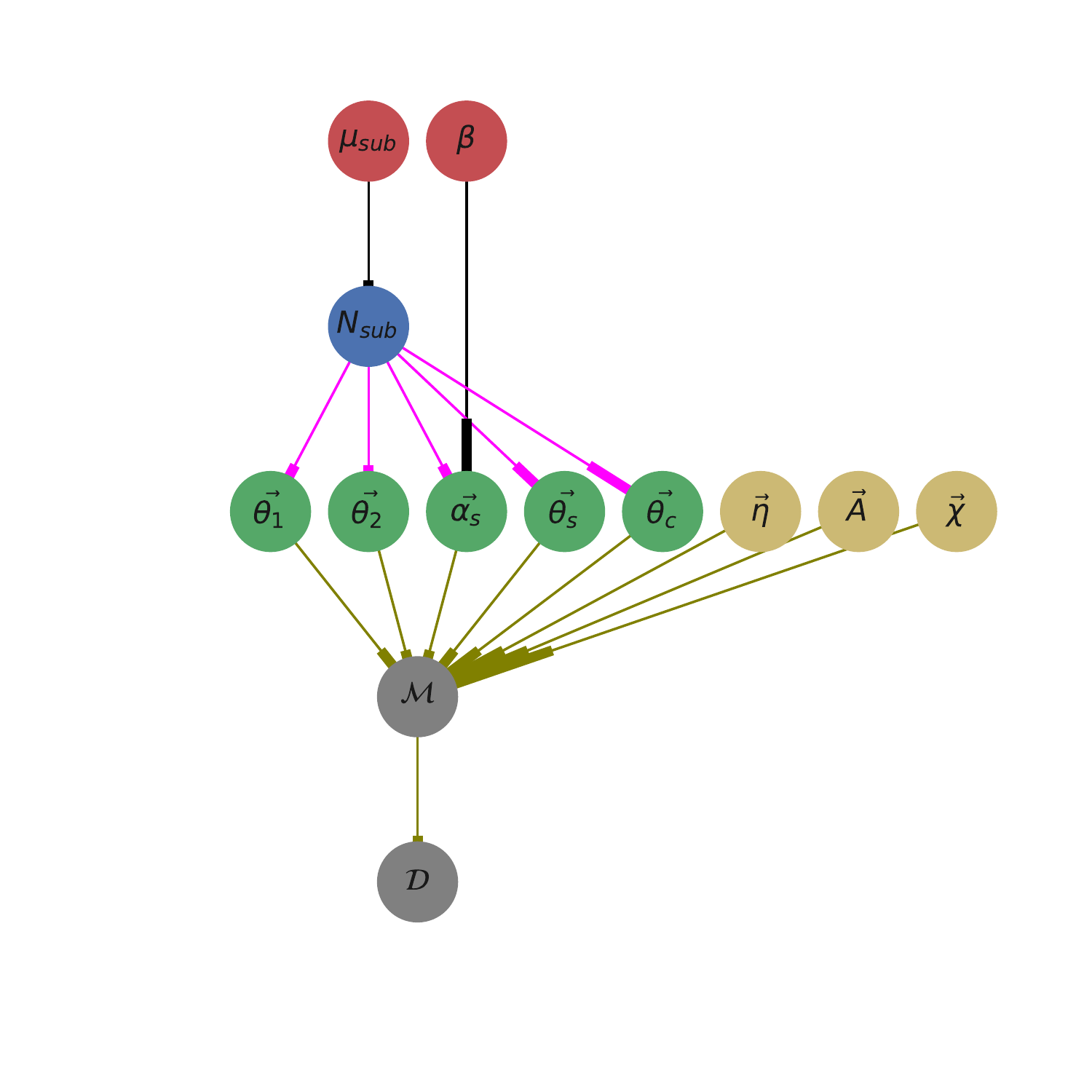}
    \caption{The probabilistic graphical model of our \texttt{PCAT} lens metamodel. Each colored node in the network corresponds either to a single parameter or a set of parameters (when vectorized) in the metamodel. Here, $\mathcal{M}$ denotes the forward-modeled photon count maps and $\mathcal{D}$ stands for the observed photon count maps. Nodes and edges are colored depending on the type of parameter they represent and their conditional dependences, respectively. See Section \ref{sect:modlsumm} and Table \ref{tabl:modl} for further details.}
    \label{figr:grap}
\end{figure*}

The hierarchical nature of probabilistic cataloging can be seen in Figure \ref{figr:grap}, where we show the probabilistic graphical model associated with the lens model. The nodes in the graph represent random variables with a certain probability density, and directed lines indicate the conditional relation between random variables. Green nodes correspond to subhalo parameters, whose multiplicity, $N_{\rm{sub}}$, is another (discrete) parameter shown with blue. Red nodes denote the hyperparameters, which control the distribution of subhalo parameters such as multiplicity and distribution of Einstein radii. For instance, the number of subhalos, $N_{\rm{sub}}$, is assumed to be a Poisson realization of an underlying mean number of subhalos, $\mu_{\rm{sub}}$. Similarly, the deflection strengths, $\alpha_{\rm{s},\textit{n}}$, are distributed as a power-law with slope $\beta$.

The yellow nodes are the fixed-dimensional parameters, which affect the model image. Here, $\vec{A}$ collectively represents the set of parameters that control the background emission. In this work, we assume that the background is spatially uniform and data (hence the model) is in a single energy band. Therefore, $\vec{A}$ is reduced to a single parameter, $A$, which is the amplitude of spatially uniform emission in units of erg/cm$^2$/s/sr. We collectively denote all the macro lens parameters with $\vec{\chi}$. We use $\vec{\eta}$ to denote the set of parameters that characterize the PSF of the instrument 

\section{Probabilistic Cataloging}
\label{sect:pcat}

In this section, we give an overview of probabilistic cataloging and refer the reader to \citep{Brewer2012, Daylan2016, Portillo2017} for details. Probabilistic cataloging is a transdimensional sampling scheme, where an a priori unknown number of elements, each with certain features, are inferred from an observed data set, e.g., a series of images, through evaluation of the Poisson likelihood of a model given the data. We use the \texttt{python2.7} implementation of the \texttt{PCAT} framework, which is publicly available at \url{https://github.com/tdaylan/pcat}. \texttt{PCAT} is a parallelized Poisson mixture sampler originally designed to sample from a gamma-ray emission metamodel consistent with the Fermi-LAT data. In this work, we extend \texttt{PCAT} to evaluate lens models.

Given some Poisson-distributed count data, \texttt{PCAT} proposes reversible jumps \citep{Green1995} across members of the metamodel in the form of births, deaths, splits, and merges of transdimensional elements, which represent dark, light-deflecting subhalos in this context. Reversibility of the transdimensional proposals ensure that detailed balance is respected across the metamodel. The resulting chain contains fair samples from the posterior distribution of the metamodel given the data, which can be used to compare models or constrain their parameters.

When sampling from the posterior probability density of the metamodel, we assume that a photon-counting experiment, e.g., a CCD, measures impinging photon counts in spatial pixels and spectral bands. We denote the observed number of photons in energy band $i$ and pixel $j$ by $k_{ij}^{\rm d}$. Given an observed image, ideally our aim would ideally be to infer the underlying true model, $k_{ij}^{\rm true}$, such that $k_{ij}^{\rm d}$ is a Poisson realization of $k_{ij}^{\rm true}$. In practice, however, the true model can be highly complex. Therefore, we approximate it using a parametric metamodel, $k_{ij}^{\rm m}$, that predicts the \emph{mean} photon count in energy band $i$ and pixel $j$. The observed photon count map can therefore be written as a Poisson realization of the photon count map predicted by the approximate model, yielding the log-likelihood

\begin{align}
\begin{split}
    \ln P(\mathcal{D}|\mathcal{M}) &= \sum_{ij} \ln P(k_{ij}^{\rm d}|k_{ij}^{\rm m}) \\
    &= \sum_{ij} k_{ij}^{\rm d} \ln k_{ij}^{\rm m} - k_{ij}^{\rm m} - \ln k_{ij}^{\rm d} !,
    \label{equa:llik}
\end{split}
\end{align}
where $\mathcal{D}$ represents the set of observed photon count maps and $\mathcal{M}$ denotes those predicted by the metamodel. The photon count map predicted by the metamodel in energy band $i$ and pixel $j$, $k_{ij}^{\rm m}$, is then the model flux convolved with the exposure of the detector toward a particular direction on the sky, ($\theta_1, \theta_2$), at energy $E$, $\epsilon(E, \theta_1, \theta_2)$, and the transmission efficiency of the optical filter, $T(E)$,
\begin{multline}
    k_{ij}^{\rm m} = \\ \iiint f_{\textrm{m}}(E, \theta_1, \theta_2) \epsilon(E, \theta_1, \theta_2) T(E) \dd{E} \dd{\theta_1} \dd{\theta_2},
\end{multline}
where $f_{\textrm{m}}(E, \theta_1, \theta_2)$ is the image predicted by the metamodel, i.e., number of photons per area, time, energy, and solid angle. In other words, we project the prediction of the lens metamodel onto the space where data are measured. In this space of pixels and energy bands, measurement uncertainties are simply Poissonian, which makes error propagation straightforward. Once $k_{ij}^{\rm m}$ is calculated over all pixels and energy bands, the Poisson log-likelihood of observing the image, given the data, is summed over all pixels and energy bands as in Equation \eqref{equa:llik}, which yields the test statistics that can be used to compare models. Note that the term $\ln k_{ij}^{\rm d}$ is fixed for a given data set, and is neglected when comparing the likelihood of different models for the same data set.

Because inference is based directly on the observed image and involves a fairly large model space, probabilistic cataloging is a time-consuming method. However, parallel computing and fast evaluation of the log-likelihood significantly reduces the required execution times. For the results presented in this paper, 10 chains with $2 \times 10^6$ samples each were collected in parallel over 8 hr (72 CPU-hours). By evaluating the likelihood in a compiled language, the execution speed can be increased by a factor of $\sim 5$ in the future to allow sampling simultaneously from the catalog space of multiple strong lens systems in the future.

\section{Results}
\label{sect:resu}

In order to validate the inference performance of \texttt{PCAT}, we run it on mock (simulated) data. We generate these mock count maps as Poisson realizations of the photon count maps predicted by forward-modeling a metamodel, which we refer to as the \emph{true} metamodel. Because we know the parameters of the true metamodel, we can compare the posterior probability distributions sampled by \texttt{PCAT} with the underlying true parameters. This approach yields full control over systematic errors by allowing us, for instance, to fit a data set that has been drawn from a different metamodel.

\subsection{Mock Data Set}
\label{sect:mock}

When simulating an image, we assume that the data are taken using the WFC3/UVIS detector on the Hubble Space Telescope (HST). This choice is only intended to provide a concrete example, because probabilistic cataloging can be applied to any photometric data set such as ground-based, high-resolution images that use adaptive optics. We further assume that the F814W filter is used to collect photons. For this HST band, we multiply differential flux predicted by the metamodel, i.e., in units of erg cm$^{-2}$ s$^{-1}$ \AA$^{-1}$ sr$^{-1}$, by $6.1\times10^{18}$ erg$^{-1}$ cm$^2$ s \AA \citep{Ryan2016}, as well as the pixel area in order to obtain the number of photons predicted per pixel. We assume that the photometric data are collected in a single exposure and that no drizzling is applied. Furthermore, we fix the observation time to 1000 seconds, which is roughly equivalent to half an orbit of HST. Choosing the mock galaxy brightnesses low enough allows one to ensure that the CCD does not sature over the selected exposure time. Finally, defining SNR as the ratio of the lensed surface brightness with the square root of the total surface brightness, our nominal mock data set has a maximum per-pixel SNR of $9$, which quickly decays away from the multiple images of the background source.

We sample the parameters of the true metamodel randomly from its prior, unless stated otherwise in Table \ref{tabl:modl}. In particular, we choose the minimum of the true distribution of $\alpha_{{\rm s},n}$ to be 0.003 arcsec, and fix the number of true subhalos to 25. These two choices are made so that the resulting true subhalo mass fraction, averaged within 0.1 arcsec of the critical curve (see the discussion on the subhalo mass fraction in Section \ref{sect:nomi}) is $\sim2\%$ \citep{Gao2011}. Furthermore, we assign deterministic values to some parameters as indicated under the \emph{true} column of the table. The letter combination RC under this column indicates that the associated parameters (galaxy coordinates) are drawn from Gaussians at the center of the image with a standard deviation equal to the pixel size (0.04 arcsec). Likewise, R denotes that the parameter is drawn randomly from the prior. Once the true metamodel parameters are determined, the predicted image is calculated and a Poisson realization of the map is drawn to obtain the mock photon count map.

Note that the true subhalos in this mock data set are mostly low-significance subhalos. Out of the 25 true subhalos, only 3 improve the goodness of fit, $\Delta \log P(D|M)$, by more than 35 when added to the best-fitting model, meaning that the rest are below formal detection ($5\sigma$ for five degrees of freedom).

\subsection{Nominal Results}
\label{sect:nomi}

\begin{figure*}
    \centering
    \includegraphics[width=0.8\textwidth, trim=1cm 2cm 0cm 1cm, clip]{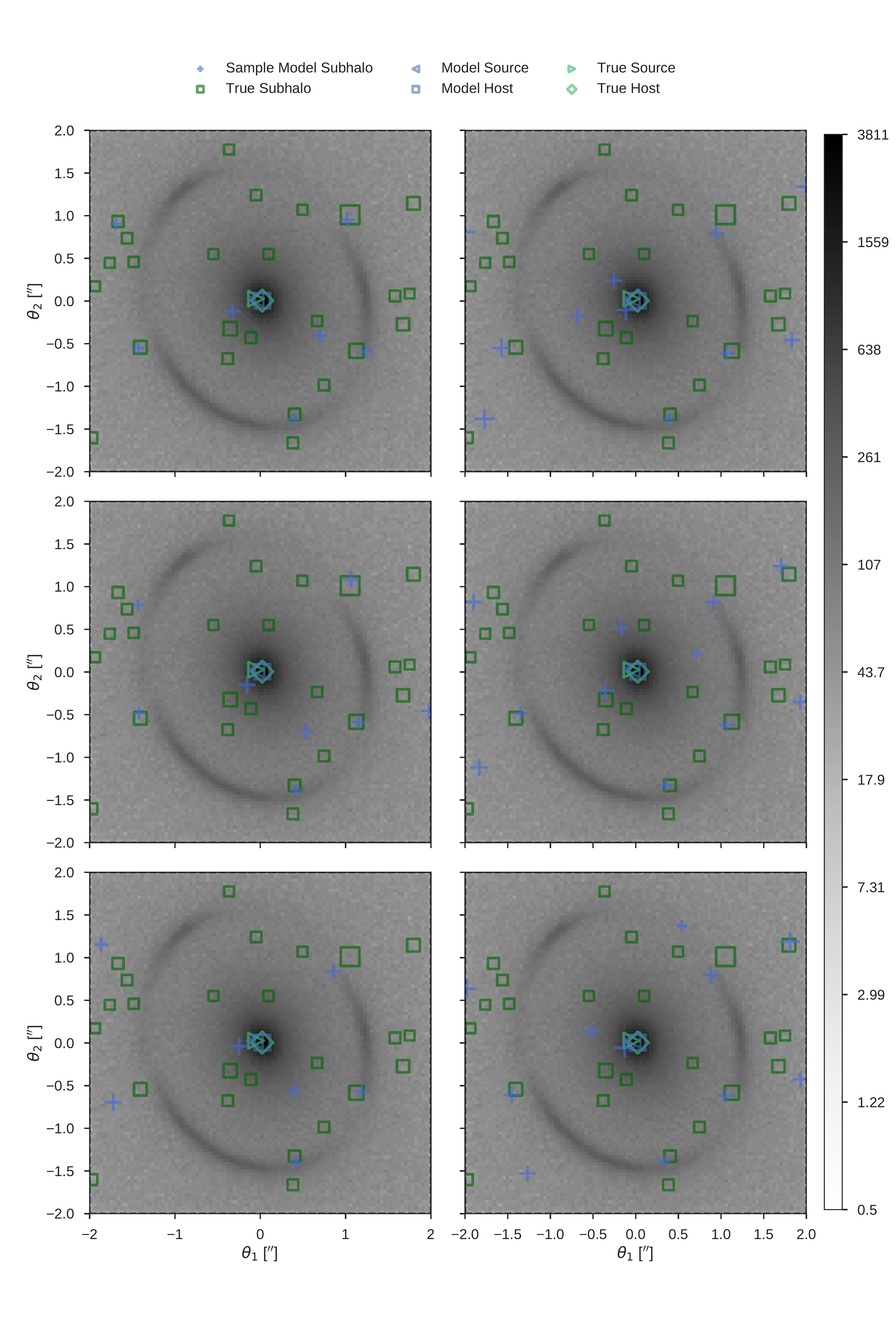}
    \caption{Fair samples from the catalog space of subhalos. The gray scale shows the number of photons per pixel using a $\log$ stretch. Superposed on the mock image, the panels show 6 out of 1000 fair samples from the subhalo catalog space, using blue crosses. The posterior samples shown in the left column have a \emph{parsimony} prior imposed on them, which effectively reduces the number of subhalos (hence the metamodel complexity). The samples on the right have the default (i.e., uniform P($\log \mu$)) prior. Also superposed on all panels is the true catalog of subhalos, shown with green squares. The sizes of the markers are proportional to the square root of the deflection strengths of the subhalos. The green diamond and blue square denote the position of the true and sample host galaxy, respectively. Similarly, the green right triangle and the blue left triangle show the position of the true and sample background source, respectively. Note that the macro lens model is allowed to change from panel to panel.}
    \label{figr:mosapop0ene0evtt0}
\end{figure*}

We begin by showing, in Figure \ref{figr:mosapop0ene0evtt0} one realization of the mock photon count map (repeated across panels) and six fair samples from the ensemble of subhalo catalogs, represented with blue pluses. The true catalog of subhalos is indicated by the green markers.

The multiple images in the simulated data set are formed by a background galaxy near the fold caustic of an elliptical foreground galaxy. The green right triangle and the blue left triangle show the position of the true and model background source, respectively. Similarly, the green diamond and the blue filled square indicate the positions of the true and model host galaxy on the lens plane, respectively. The positions of the true and model galaxies are overlapping in all samples, although the positions of the blue markers have unnoticeable variations across samples due to the rather small uncertainties in the posterior distributions of the source and host galaxy positions.

\paragraph{\textbf{Macro lens model}}

In this work, we focus on the inference of subhalo properties in the lens metamodel. Nevertheless, subhalos have only a perturbative effect on the deflection field, which is otherwise dominated by the host galaxy. Therefore, any bias in the host galaxy parameters, as well as other macro lens parameters -- such as those of the background galaxy and the external shear -- can potentially leak into constraints on the subhalo properties. We therefore first discuss the posterior distribution of the macro lens model parameters.

In general, we find that the best-constrained macro lens parameters are the positions of the host and source galaxies as well as the Einstein radius of the host galaxy. They generally have posterior 68$\%$ credible intervals that are $6\times10^{-4}$ arcsec, $5\times10^{-3}$ arcsec, and  $0.01$ arcsec wide, respectively. However, we also find that the Einstein radius of the host halo correlates with its ellipticity, the amplitude and orientation angle of which are constrained inside 68\% credible intervals that are 0.01 and 2$^\circ$ wide, respectively. This anti-correlation can be attributed to the fact that the critical curve is constrained more strongly along one axis of symmetry than the other.

\begin{figure}
    \includegraphics[width=0.45\textwidth]{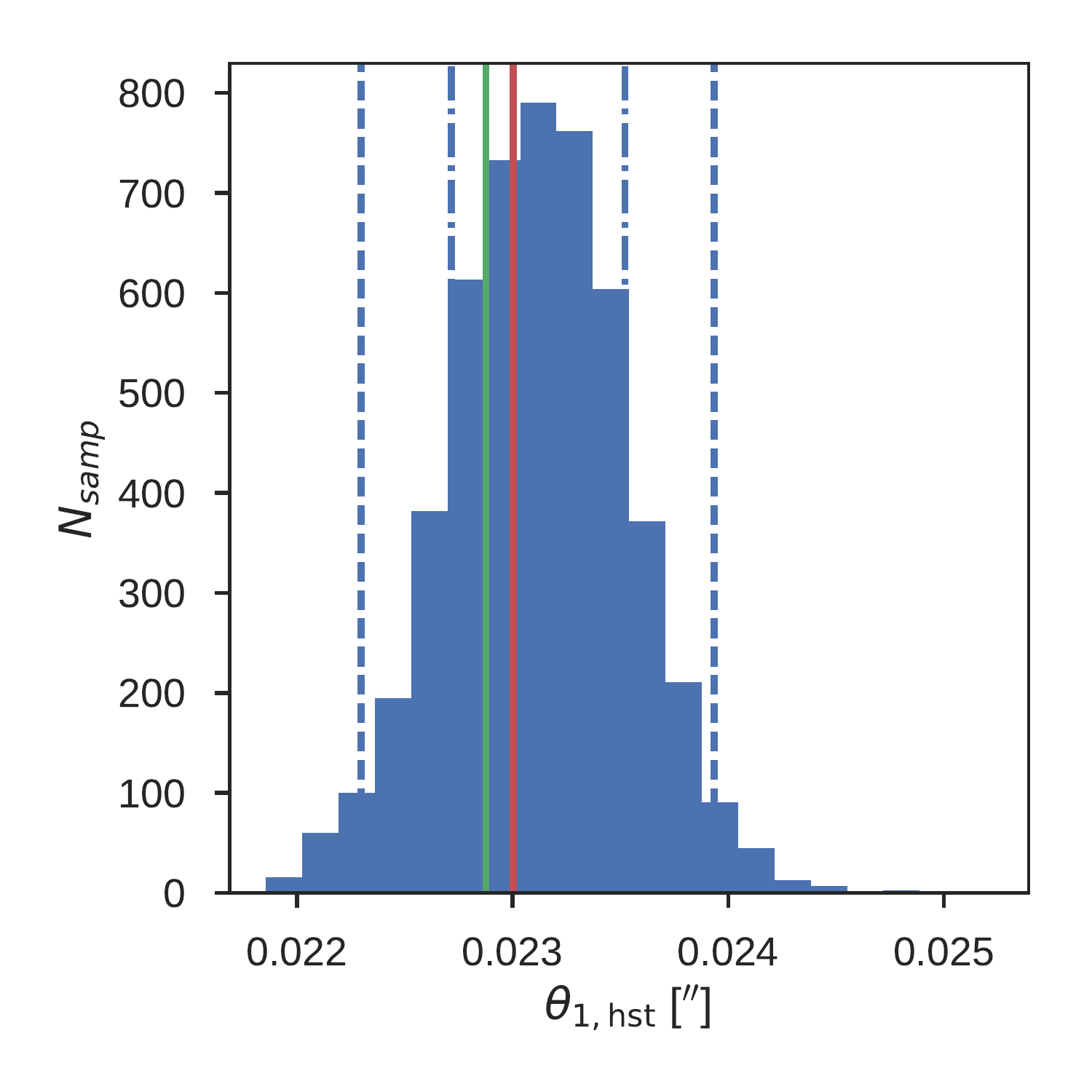}
    \caption{Histogram of samples from the posterior distribution of the horizontal position of the host halo shown with the blue histogram. The green vertical line indicates the true value of the parameter, whereas the red vertical line shows the value of the parameter in the maximum likelihood sample. The dotted-dashed and dashed vertical lines denote the following percentiles: 2.5, 16, 84, and 97.5.}
    \label{figr:lgalhost_hist}
\end{figure}

As an example, Figure \ref{figr:lgalhost_hist} shows the distribution of posterior samples on the horizontal position of the host halo. The green vertical line indicates the true value of the parameter, whereas the red vertical line shows the value of the parameter in the maximum likelihood sample. The dashed and dotted-dashed vertical lines show the 95\% and 68\% uncertainties around the median.

The host and source galaxy fluxes, on the other hand, are constrained at the $\sim 2\%$ and $\sim 10\%$ level, respectively. Specifically, the fact that we impose the background galaxy to have a S\'ersic profile of known index reduces the uncertainty on it. If the source plane emission had been allowed to vary from pixel to pixel, subject to some regularization, the uncertainties on the background emission would be even larger. Nevertheless, because we are mainly interested in modeling the foreground mass distribution, the background emission is eventually marginalized out, which reflects the uncertainty of the marginal posterior probability distribution of the subhalo properties. The choice of S\'ersic profile also introduces a strong covariance between the luminosity and half-light radius of the galaxies, which can have correlation coefficients as large as 0.9. 

The external shear is the least strongly constrained ingredient of the macro lens model, with typical uncertainties of $\sim 10\%$ on the magnitude of the shear field and $\sim 20^\circ$ error on its direction. The large uncertainty in the alignment is likely due to the degeneracy of the quadrupole shear term with the overall deflection field due to subhalos. Similarly, the posterior probability distribution over the lens metamodel is marginalized over the shear field and its uncertainties are propagated to the uncertainties on the subhalo properties.

Finally, we obtain $\sim 0.1\%$ uncertainties on the isotropic background emission and recover our prior uncertainty on the PSF width in the posterior distribution. The latter allows one to marginalize over uncertainties that are due to our lack of perfect knowledge of the PSF, and is not intended to gain information about the PSF using the image.

\paragraph{\textbf{Subhalo mass function}}

\begin{figure}
    \includegraphics[width=0.45\textwidth]{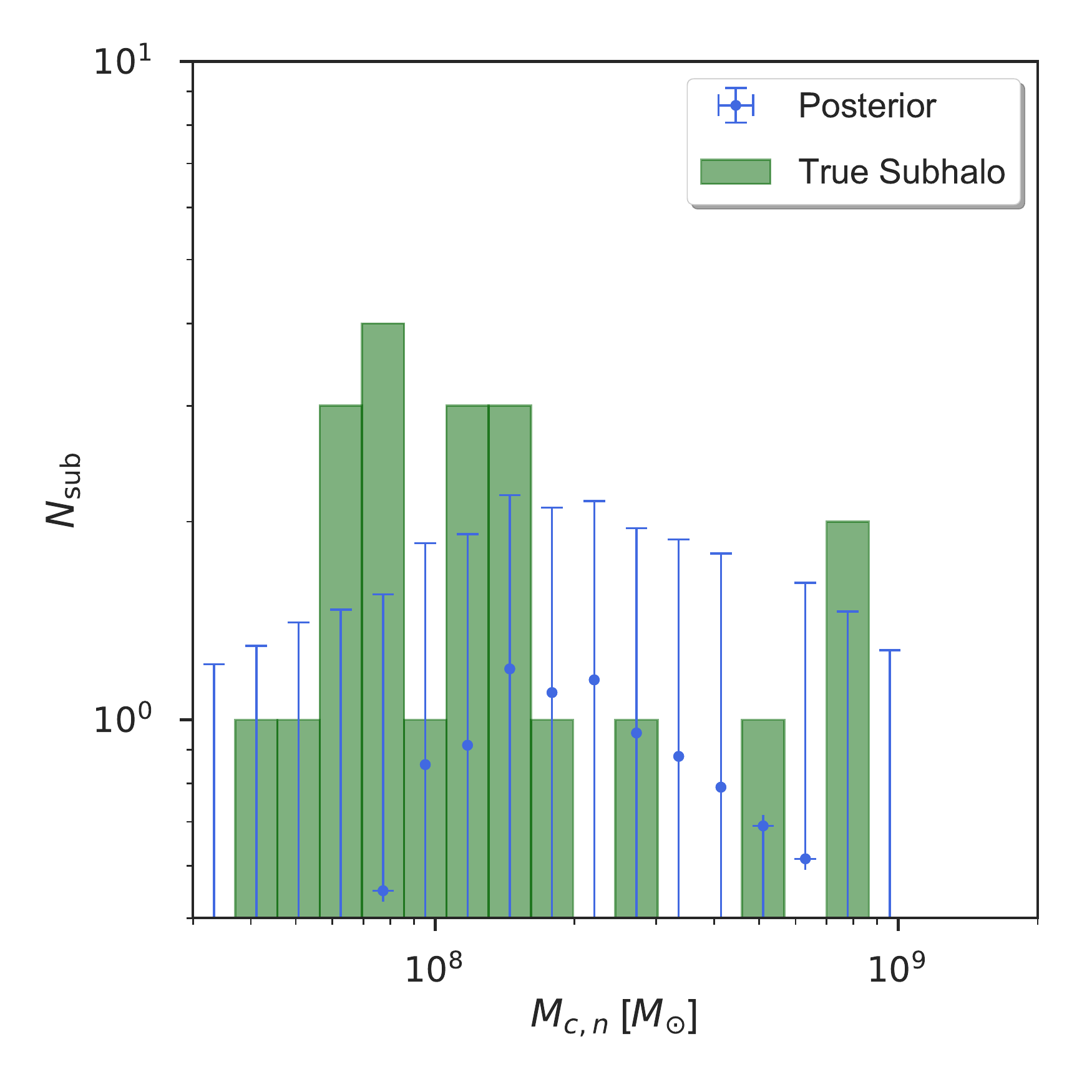}
    \caption{Histogram of the tidally truncated subhalo masses in the mock model (green histogram) and the posterior distribution (blue error bars). The central point and the error bars show the 16th, 50th, and 84th percentiles of the histogram samples from the posterior, respectively. The posterior distribution agrees with the true distribution in a setting where the majority of the subhalos are below the detection threshold. We stress that the inferred posterior depends on our choice of $\alpha_{\rm{s,min}}$, i.e., the posterior would extend to lower masses (and be more prior-dominated) if we lowered $\alpha_{\rm{s,min}}$. Therefore, the main inference of interest here is the \emph{normalization} of the posterior subhalo mass distribution, \emph{given} a choice of $\alpha_{\rm{s,min}}$ and a fairly broad Gaussian prior on the log-slope, $\beta$.}
    \label{figr:pdfnhistmcutpop0reg0nomi}
\end{figure}

We present the posterior truncated mass distribution of the subhalos in Figure \ref{figr:pdfnhistmcutpop0reg0nomi}. The figure shows the histogram of the truncated masses of the true subhalos in green. Overplotted with blue error bars, we show the median histogram of the truncated masses of the subhalos in the posterior samples. These blue points with error bars are obtained by first calculating the histogram of the truncated masses for each posterior sample from the metamodel. The 16th, 50th, and 84th percentiles in each bin are then plotted as the lower cap, central point, and the upper cap of the error bar, respectively.

It may initially seem surprising that even the highest-mass bins that contain a true subhalo may have a median posterior at zero, or that the posterior uncertainties go above zero when there is no true subhalo. This is a natural result of accounting for transdimensional (across-model) covariances in the problem. When births and deaths of subhalos are allowed, a true subhalo can sometimes be fitted with multiple and lower-mass subhalos. Similarly, multiple true subhalos can be fitted with a single more massive model subhalo. It is important to emphasize that such transdimensional covariances exist even for true subhalos with high statistical significance. In other words, even though the log-likelihood difference between two models with and without a given subhalo may be well above 35, it may be as low as a few when comparing two models where the subhalo exists as a single, high-mass clump and as two, smaller-mass clumps. Furthermore, it is also true that a high-mass subhalo can be far from the multiple images, causing the posterior subhalo mass function to be uninformed of its existence.

\begin{figure}
    \includegraphics[width=0.45\textwidth]{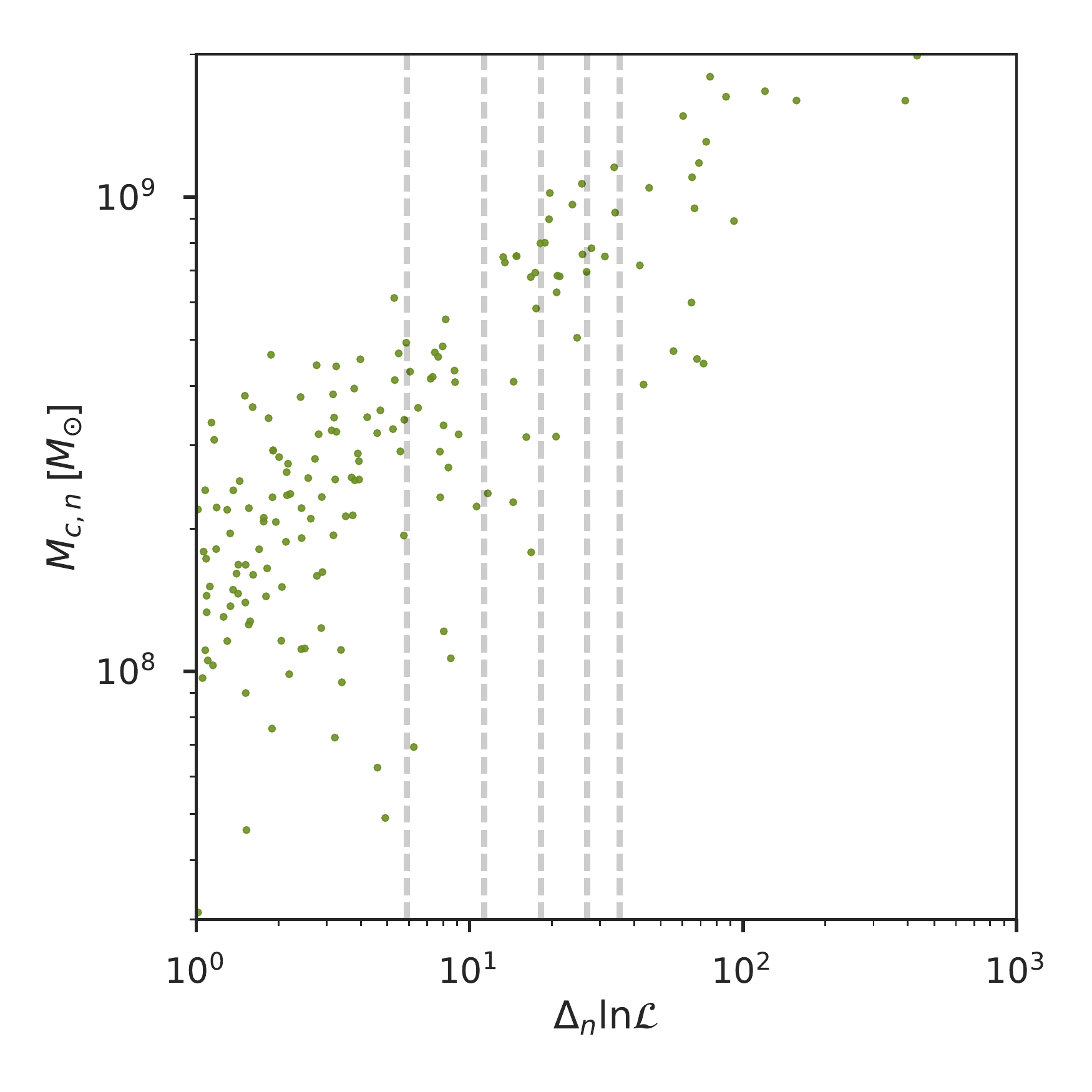}
    \caption{Truncated mass vs. significance (i.e., maximum log-likelihood difference of adding the subhalo to the model) of mock subhalos in a different simulation with a higher number of subhalos. The vertical lines indicate 1$\sigma$ through 5$\sigma$ for five degrees of freedom. For this simulation, subhalos fall below 1$\sigma$ at a mean truncated mass of $2 \times 10^8 M_\odot$.}
    \label{figr:scatdeltllikmcut}
\end{figure}

We note that the roll-off at small masses is due to the choice of the minimum deflection strength allowed in the model, $\alpha_{\rm{s,min}}$. If this value is decreased, the posterior probability density extends to lower masses. Therefore, the constraints derived on the subhalo mass function must be quoted for a given $\alpha_{\rm{s,min}}$. Given our choice of the mock data set and fiducial redshifts, our posterior subhalo mass function becomes prior-driven below $\sim 10^8 M_\odot$. This is illustrated in Figure \ref{figr:scatdeltllikmcut}, where we plot the truncated mass and significance of subhalos in a different mock data set with a higher number of subhalos.

As for the deflection profile parameters $\theta_{{\rm s},n}$ and $\theta_{{\rm c},n}$, we generally find that the projected cutoff radius is better constrained than the projected scale radius. In fact, WFC3's pixel size of 0.04 arcsec is comparable to the projected scale radius of an NFW subhalo with $r_{{\rm s},n}=200$ pc at the fiducial redshift of $z=0.2$. This implies that current optical photometry is insensitive to the scale radii of most subhalos. However, more massive subhalos close to the multiple images can still be distinguished by their projected scale radii, which motivates our choice to make $\theta_{{\rm s},n}$ subject to inference.

Note that the mass of a subhalo does not have a one-to-one correlation with its statistical significance. In a photometric inference problem, how the flux of a given point source compares to the background emission inside the full width at half maximum (FWHM) of the PSF, largely determines its statistical significance \citep{Portillo2017}. Therefore, we expect that isolated, bright light sources will be, on average, to be more statistically significant than dimmer ones. However, how strongly a subhalo can be constrained in the lensing problem depends on its mass, as well as how its deflection field is oriented with respect to the lensed emission. In other words, because the deflection profile of a subhalo drops beyond its projected scale radius, a subhalo needs to be close to an already lensed emission on the image plane \emph{as well as} have a sufficiently large mass, in order to have non-negligible effect on the log-likelihood.

The leading-order contribution of a given subhalo with index $n$ to the model image can be obtained by Taylor-expanding Equation \eqref{equa:lensflux}. Recasting this equation in terms of the photon count map lensed by the host halo, external shear, and all subhalos except the $n$th subhalo, $\bar{k}_m$, we obtain the lensed count map

\begin{align}
    \tilde{k}_m(\theta_1, \theta_2) = \bar{k}_m(\theta_1, \theta_2) + \vec{\alpha}_n \cdot \vec{\nabla} \bar{k}_m(\theta_1, \theta_2) + \mathcal{O}(\vec{\alpha}_n^2).
\end{align}
The second term is negative over some regions of the lensed emission, and positive in other regions. Instead, $\Big\langle \lvert\vec{\alpha}_n \cdot \vec{\nabla} k_m\Big \rvert \rangle$ better encapsulates the level of perturbation that a given subhalo introduces to the image already lensed by all other mass components. Therefore, it is a relatively more accurate estimator of the statistical significance of a given subhalo compared to $M_{\rm{c},n}$. We refer to this quantity as the subhalo \emph{relevance}. Subhalos with low relevance, although possibly real, cannot be constrained using the observed photometric data. An example can be seen in Figure \ref{figr:mosapop0ene0evtt0}, where the true subhalo at (1.8, 1.1) arcsec is relatively less constrained, despite having a mass above $10^8 M_\odot$.

\begin{figure}
    \includegraphics[width=0.45\textwidth]{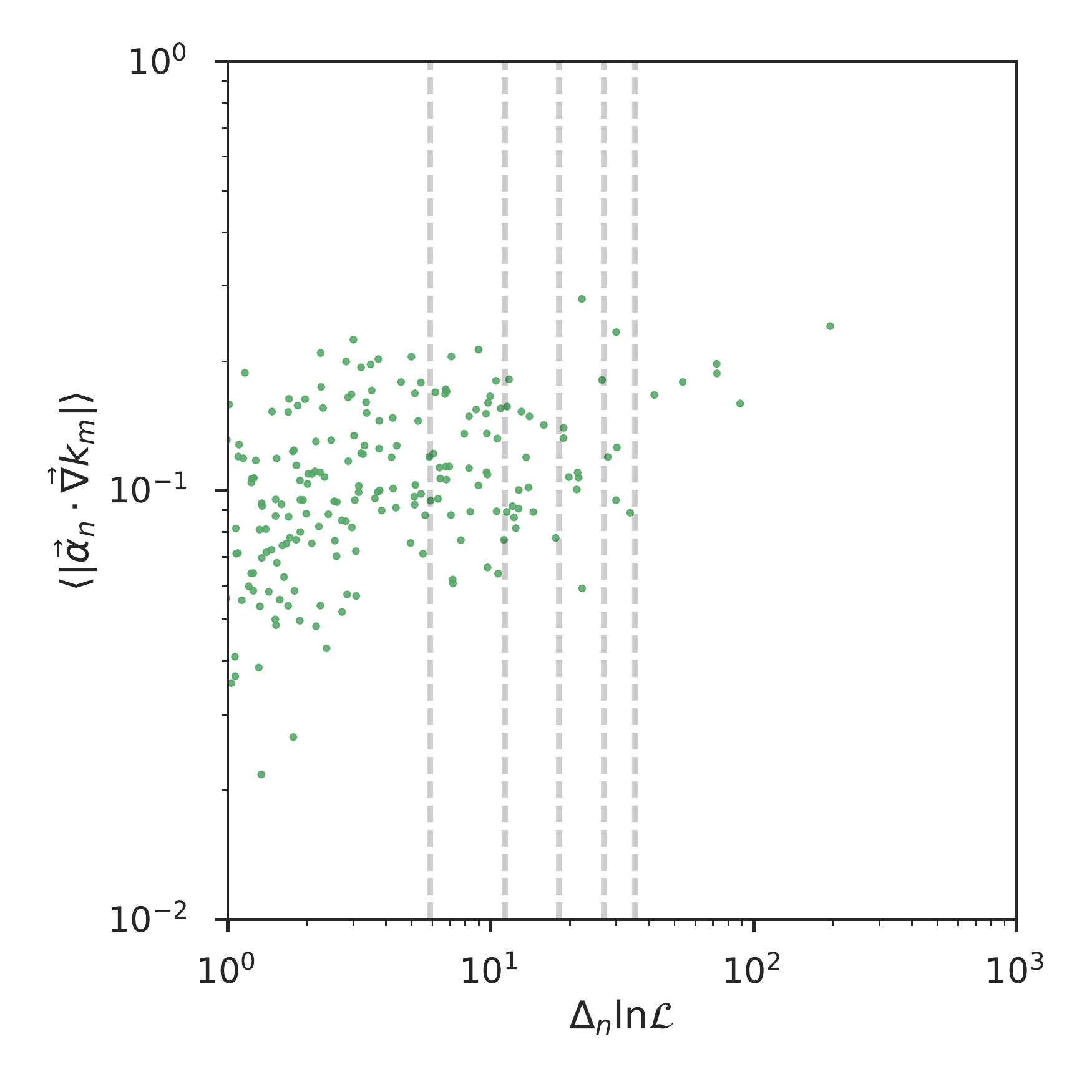}
    \caption{Correlation between the relevance of a mock subhalo population and the log-likelihood improvement in adding each subhalo to the model, $\Delta_n \ln \mathcal{L}$. Vertical dashed lines indicate $1\sigma$, $2\sigma$, $3\sigma$, $4\sigma$, and $5\sigma$ contours for five degrees of freedom. These subhalos are independently drawn and do not correspond to the true subhalos in the simulated strong lens shown in Figure \ref{figr:mosapop0ene0evtt0}.}
    \label{figr:scatdeltllikrelepop0}
\end{figure}

Figure \ref{figr:scatdeltllikrelepop0} shows the correlation between the relevance of a subhalo and the log-likelihood increase when including the subhalo into the lens model while leaving all other parameters fixed. The correlation is not one-to-one, however, due to the fact that the host galaxy and the external shear can dominate the deflection field in some regions more than others, which causes subhalos with equal relevance to have different statistical significances. Therefore, in general, significant subhalos have high relevance, but highly relevant subhalos can have very low significance.

\paragraph{\textbf{Subhalo mass fraction}}
\label{sect:fracsubh}

The shape of the subhalo mass function is poorly constrained, given the small number of strongly lensed systems analyzed so far. Nevertheless, the fraction of mass tied in subhalos in the vicinity of the critical curve of the host halo can be accurately measured for strong lenses with high SNR. Therefore, the subhalo mass fraction provides an observational summary statistic of the subhalo mass function. 

\begin{figure*}
    \includegraphics[width=0.95\textwidth]{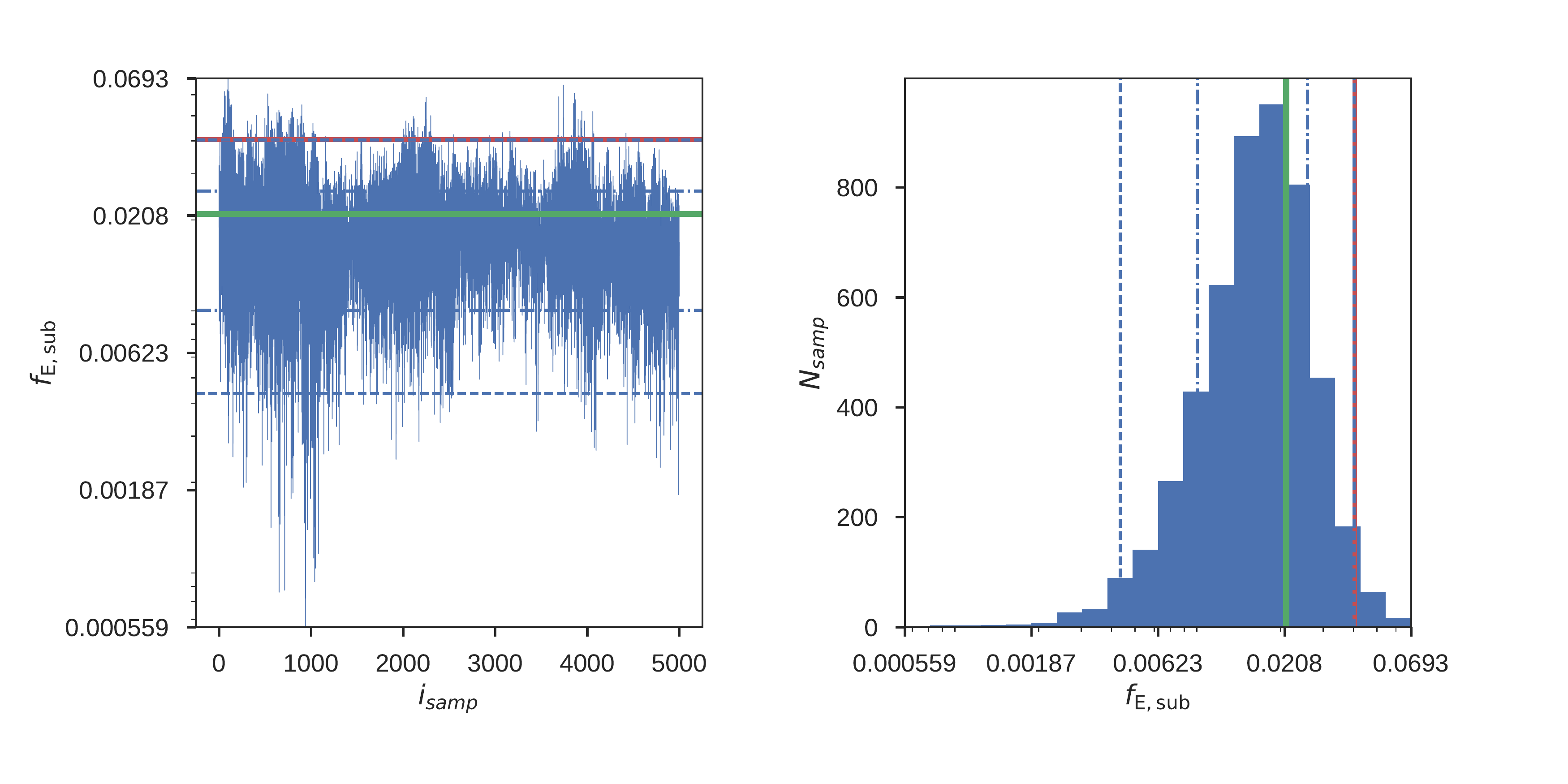}
    \caption{Histogram of samples from the posterior distribution of the subhalo mass fraction (right). The MCMC time evolution of the quantity (left). The legend is similar to that in Figure \ref{figr:lgalhost_hist}. Note that the red (maximum likelihood sample), and the dashed blue (97.5th percentile) lines happen to overlap.}
    \label{figr:fracsubhdeltbein_trac}
\end{figure*}

In order to determine the posterior fraction of mass locked in subhalos, we again first calculate it for each sample and then take posterior moments. We define the subhalo mass fraction for a given sample from the metamodel as the ratio between the mean surface mass density due to all subhalos and that due to the host halo inside an annuli centered at the host position with radius $\theta_{\rm{E,hst}}$ and thickness 0.2 arcsec,
\begin{align}
    f_{\rm{E,sub}} \equiv \dfrac{\sum_n  \displaystyle \int_{\rm{ann}} \kappa_{n}\dd^2{\theta} }{\displaystyle \int_{\rm{ann}} \kappa_{\rm{hst}} \dd^2{\theta}}.
\end{align}
where ``ann'' represents the mentioned annulus around the host halo. The resulting posterior distribution of the subhalo mass fraction at the Einstein radius is shown in Figure \ref{figr:fracsubhdeltbein_trac} along with its evolution as a function of MCMC time. Note that we do not directly impose a prior probability density on the subhalo mass fraction, $f_{\rm{E,sub}}$. The figure shows that the posterior distribution of $f_{\rm{E,sub}}$ agrees with the true value, and that it is sensitive to the existence of subhalos below detection. The green line shows the true subhalo mass fraction.

\begin{figure*}
    \centering
    \subfigure[Host halo]     {\includegraphics[width=.32\textwidth, trim=0.4cm 0cm 1.5cm 2cm, clip]{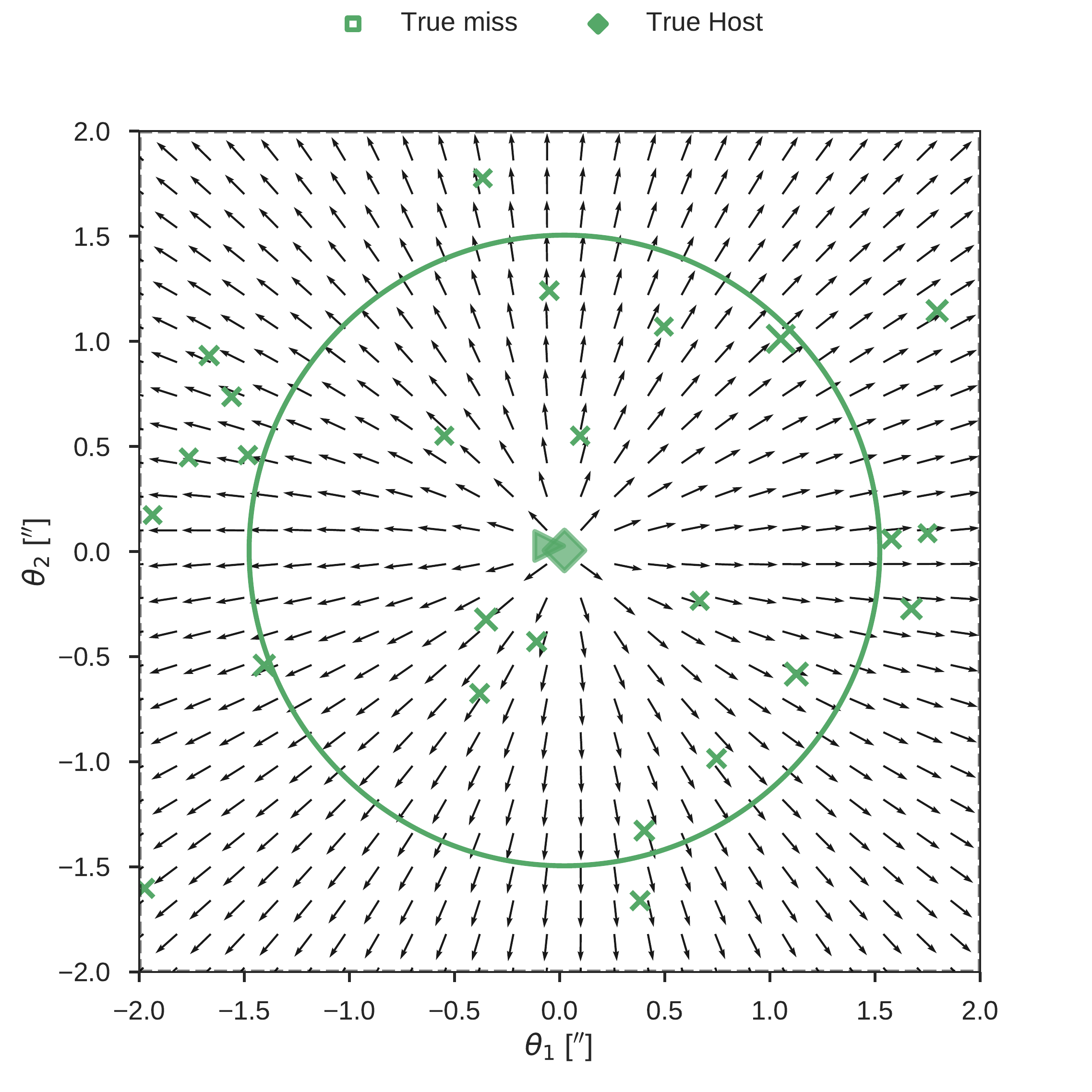}}
    \subfigure[External shear]{\includegraphics[width=.32\textwidth, trim=0.4cm 0cm 1.5cm 2cm, clip]{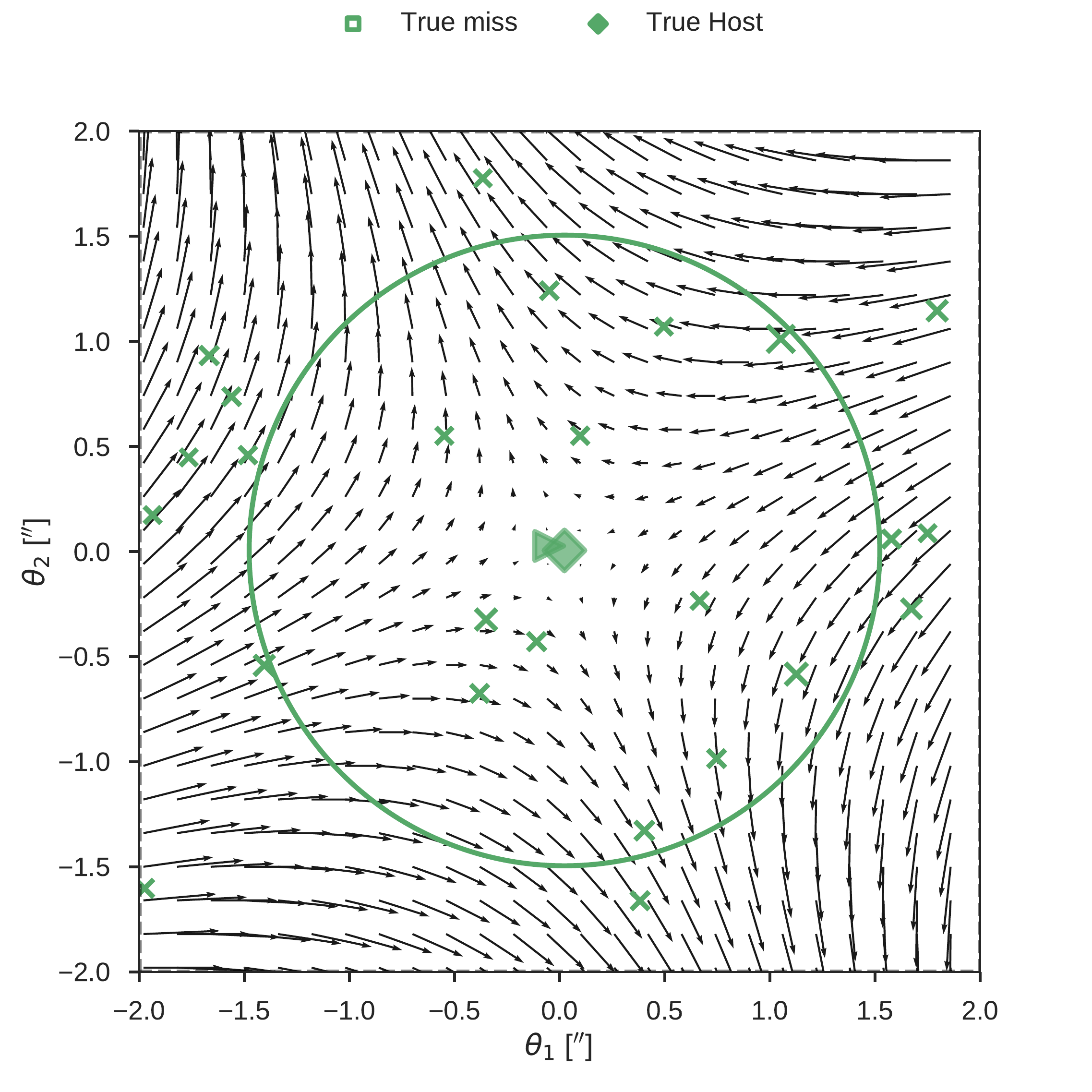}}
    \subfigure[All subhalos]  {\includegraphics[width=.32\textwidth, trim=0.4cm 0cm 1.5cm 2cm, clip]{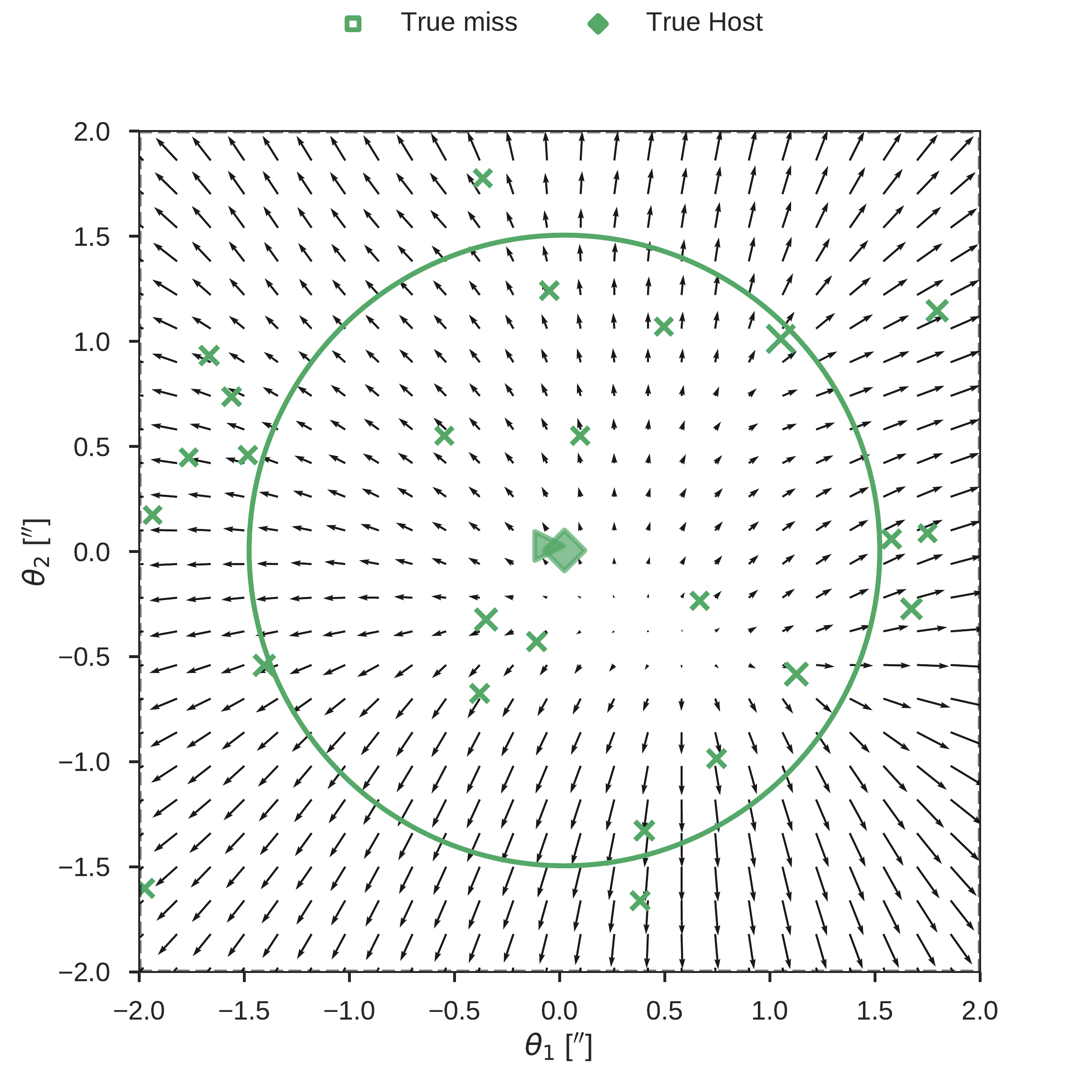}}
    \caption{Posterior median of the deflection field. The arrows indicate the local direction of the deflection field and have lengths that represent the magnitude of the deflection field. However, the arrows in the left and right panels have been scaled by 0.1 and 10, respectively, for better visibility. The positions of the true subhalos are shown with green Xs, the areas of which are proportional to the masses of the subhalos.  The radii of the green and dashed blue circles are the Einstein radii of the true and sample macro lens models, respectively. They are only drawn to guide the eye in the absence of the observed image.}
    \label{figr:postdefl}
\end{figure*}

Finally, we show the posterior deflection maps in Figure \ref{figr:postdefl}, which are obtained as follows. For each fair sample from the lens metamodel, we first calculate the deflection vector, i.e., deflection direction and magnitude, on all pixels. Then, for each pixel, we find the 50$^{th}$ percentile of the calculated deflection values, and refer to the resulting map as the posterior median deflection map. Figure \ref{figr:postdefl} then illustrates the deflection field separately for the host halo (left), external shear (center), and subhalo population (right). The radius of the green and blue dashed lines, which are also mostly overlapping, denote the circles whose radii coincide with the Einstein radii of the true and model host halos, respectively. Because the host galaxy is elliptical, however, the circles do not correspond to the critical curve -- they are only intended to guide the eye. The area of a circle is proportional to the host mass contained inside the Einstein radius. The true Einstein radius is 1.5 arcsec, yielding a mass inside the Einstein radius of 5$\times$10$^{11}$ $M_\odot$.

The posterior total deflection field is found to be constrained at the percent level across most of the image. However, this translates to $\sim$ 100\% uncertainty in the deflection field due to subhalos. The posterior deflection field due to subhalos is only constrained well near the multiple images, where the uncertainties can be as low as 20\%.

\subsection{Systematic bias in a one-subhalo lens model}

Fixed-dimensional lens modeling requires iteratively testing lens models with a fixed number of subhalos (or a fixed amount of any other form of additional model complexity) in addition to the smooth lens model. In the simplest approach, a single subhalo can be added to the smooth lens model to fit an observed photon count map. In this section, we compare the previous results from transdimensional sampling to such a case, where the lens model contains a single subhalo. To make a robust comparison, we use the same simulated photon count map as shown in Figure \ref{figr:mosapop0ene0evtt0}. Then, instead of allowing a variable number of subhalos in the lens model, we turn off transdimensional proposals in \texttt{PCAT} and fix the number of subhalos to 1. Note that the latter scheme is still Bayesian and probes all within-model covariances, but neglects across-model covariances. 

\begin{figure*}
    \subfigure[Nominal]     {\includegraphics[width=.49\textwidth, trim=0 0 0 3cm, clip]{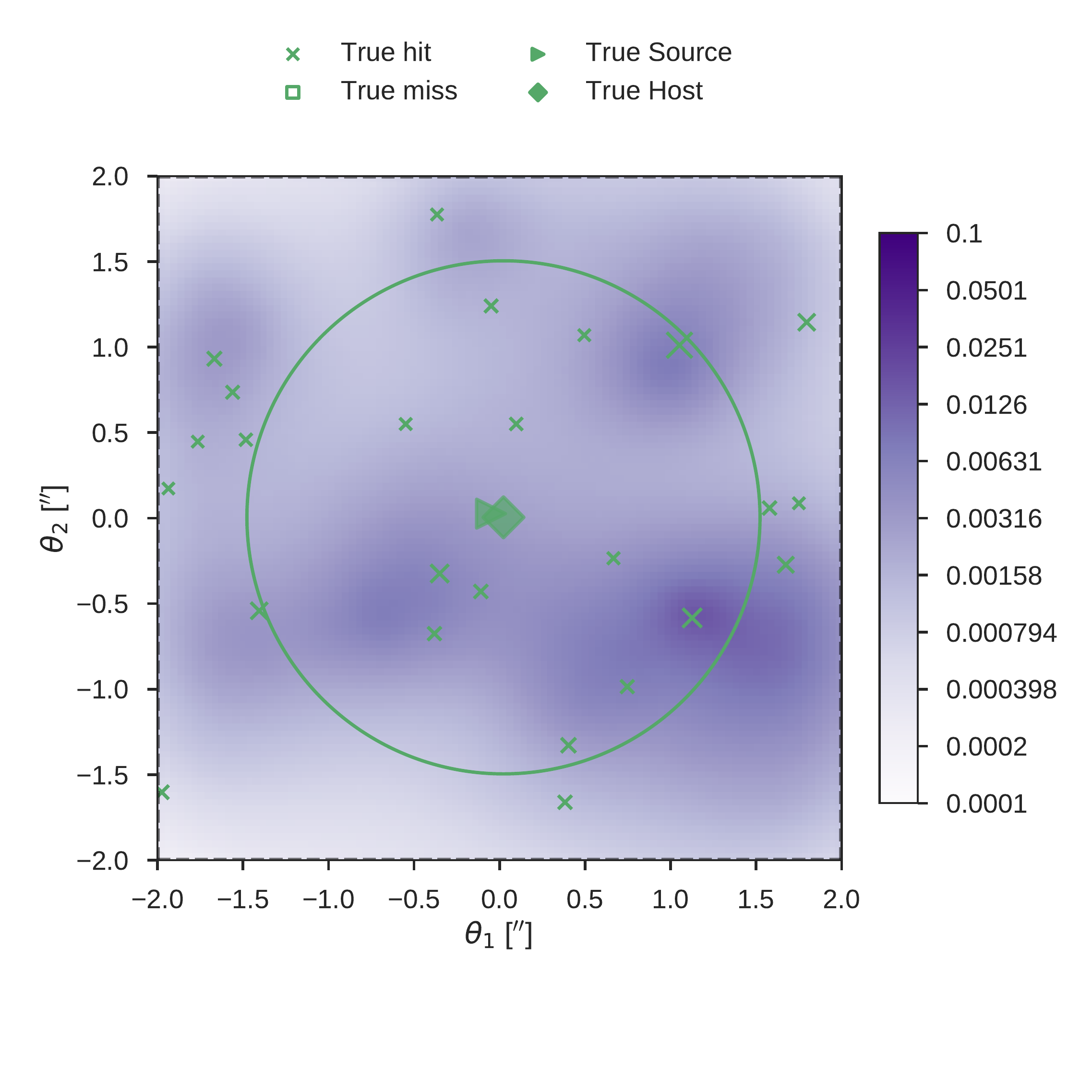}}
    \subfigure[One-subhalo] {\includegraphics[width=.49\textwidth, trim=0 0 0 3cm, clip]{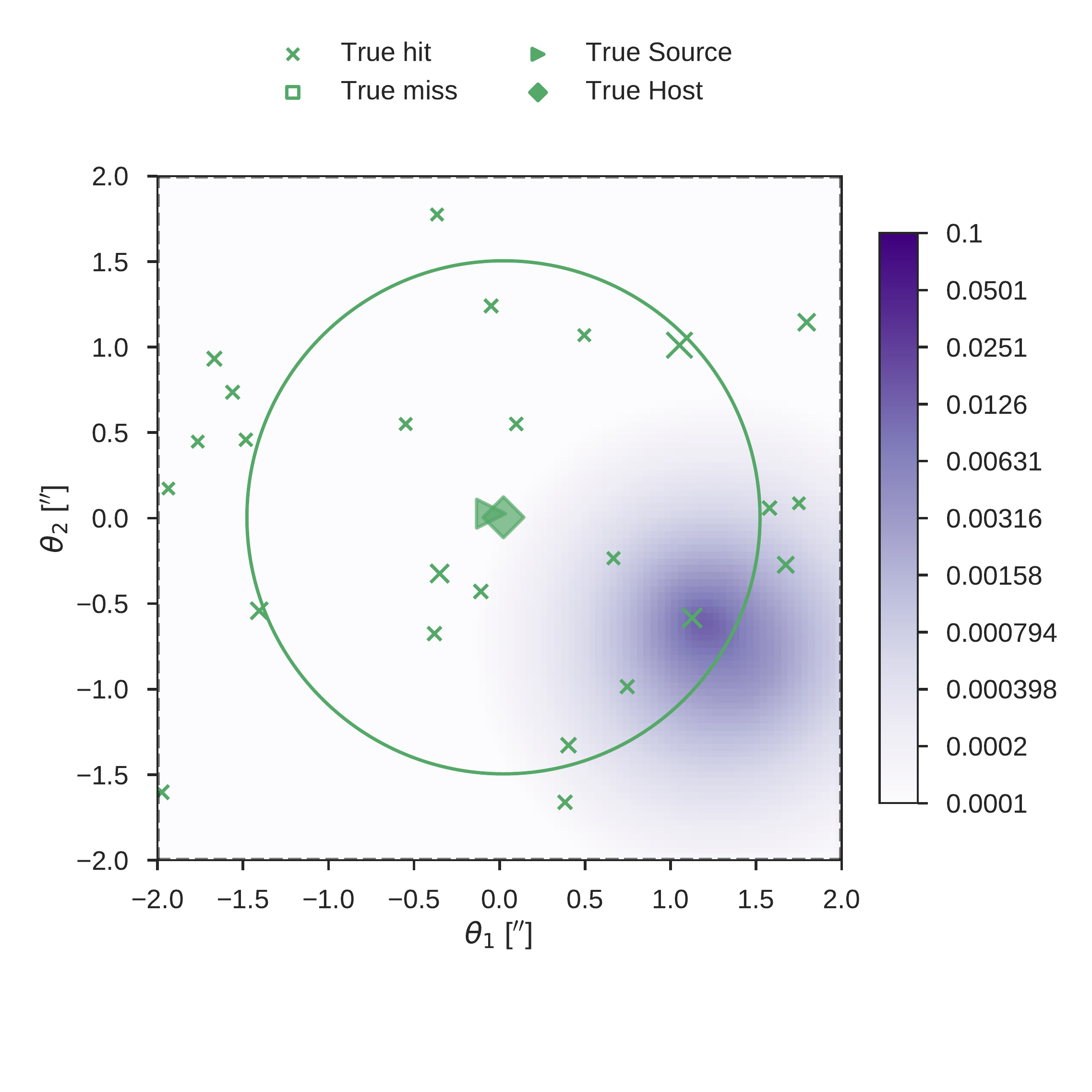}}
    \caption{Posterior median convergence map obtained from the transdimensional (left) and one-subhalo (right) inferences. Green markers and the circle have the same meaning as in Figure \ref{figr:postdefl}.}
    \label{figr:postmediconvelemevtApopA}
\end{figure*}

Figure \ref{figr:postmediconvelemevtApopA} illustrates the posterior median convergence field obtained in the transdimensional (left) and one-subhalo (right) inferences. These posterior convergence maps are obtained similarly to the posterior deflection maps. In other words, the convergence maps are calculated for each sample from the posterior by numerically differentiating the deflection field, and then the median convergence in each pixel is presented as the posterior median convergence map.

In the left panel of Figure \ref{figr:postmediconvelemevtApopA}, transdimensionality allows subhalos to be born and die across the image, probing the goodness of fit of all combinations of subhalo multiplicities, positions and deflection profiles allowed by the prior. Such configurations mostly include mild likelihood improvements below what would be called a detection threshold in conventional cataloging. The resulting posterior median convergence map reveals low-significance diffuse features in the underlying subhalo convergence field, as well as constraining statistically significant true subhalos. In contrast, the posterior median convergence map obtained from the one-subhalo inference on the right contains a single overdensity, which is also inferred by the transdimensional inference. However, the transdimensional approach also (partially) deblends subhalos in the crowded region, to the extent allowed by the information available in the data and degeneracies in the likelihood, despite the fact that most sample subhalos presented in Figure \ref{figr:mosapop0ene0evtt0} do not indicate associations with the true subhalos.

The failure of the one-subhalo inference to reveal the remaining features in the underlying true convergence field can be traced back to the fact that the absence of additional model subhalos causes the macro lens model and the single subhalo to be biased so as to minimize the residual deflection field. Without a way to probe transdimensional covariances, the sampler explores a maximum in the conditional likelihood, given a one-subhalo lens model. Note that the inference of lower-significance features in the convergence map does not imply that the additional subhalos are formally detected. In fact, the overall increase in the goodness of fit between the two approaches is only $\Delta \log P(D|M) \sim 10$. In order to claim a detection of these model subhalos at $5\sigma$, they would be required to individually improve the goodness of fit above $\Delta \log P(D|M) \sim 35$ given that they have five degrees of freedom.

\begin{figure*}
    \subfigure[Nominal]     {\includegraphics[width=.49\textwidth, trim=0 0 0 3cm, clip]{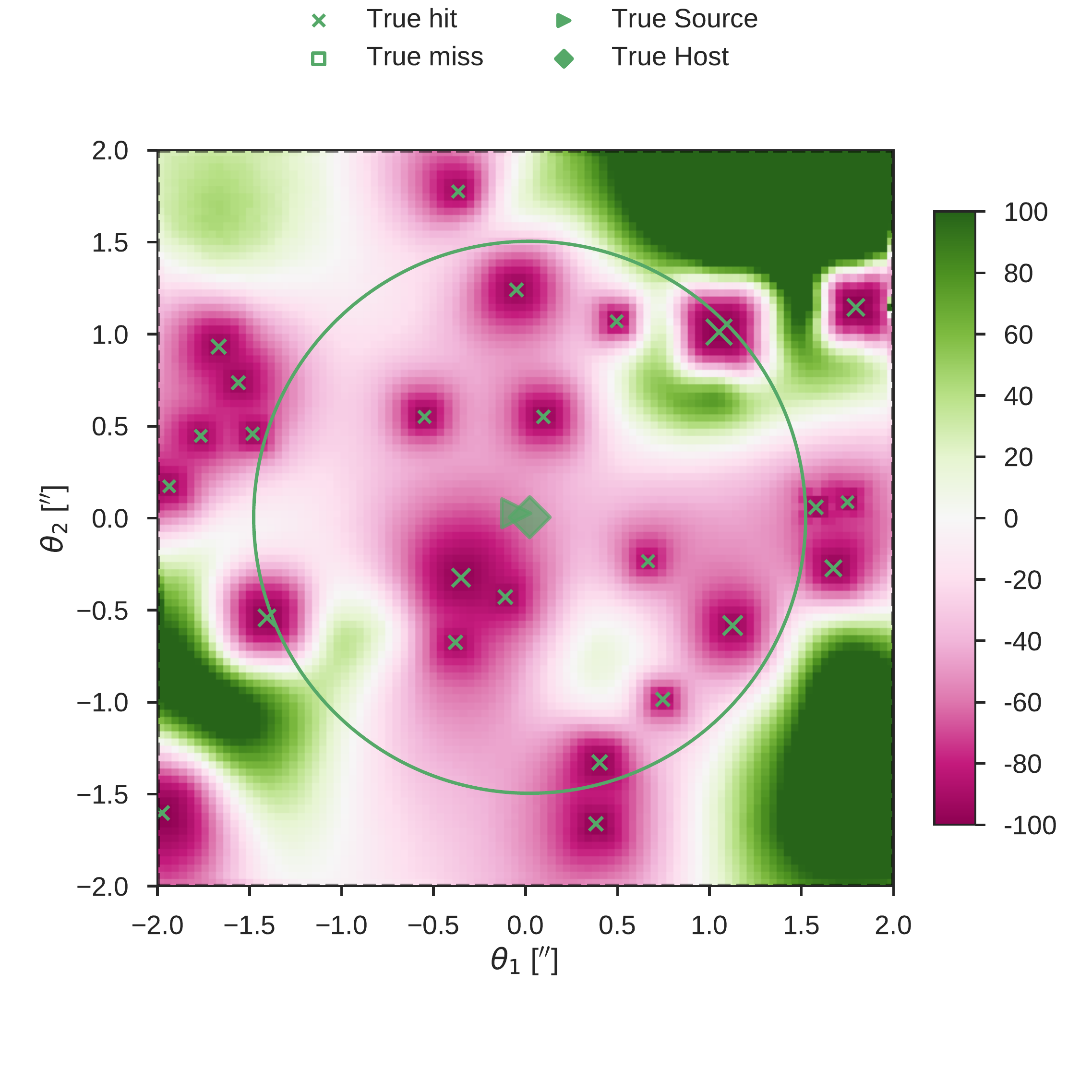}}
    \subfigure[One-subhalo] {\includegraphics[width=.49\textwidth, trim=0 0 0 3cm, clip]{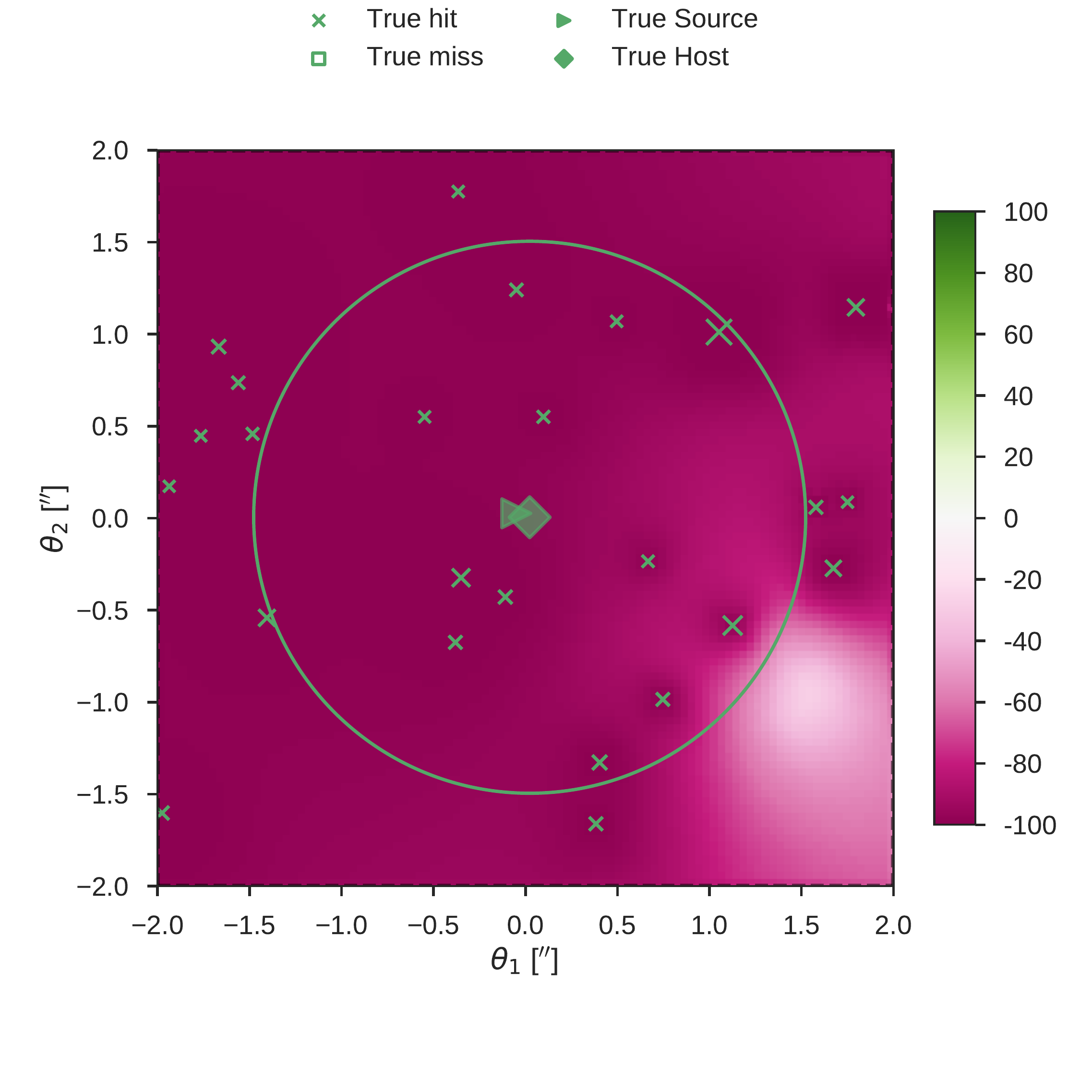}}
    \caption{Residuals between the posterior median convergence maps and the true subhalo convergence map in the transdimensional (left) and one-subhalo (right) inferences, given as percentage of the true subhalo convergence map. The divergent color scale is linear and saturated at $\pm$ 100. Green markers and the circle have the same meaning as in Figure \ref{figr:postdefl}.}
    \label{figr:postmediresiconvelempercevtApopA}
\end{figure*}

The difference between the two inference schemes becomes more evident when the posterior predictions are compared to the underlying true convergence field due to subhalos. Figure \ref{figr:postmediresiconvelempercevtApopA} illustrates the residual between the true subhalo convergence map and the posterior median convergence maps obtained in the transdimensional (left) and one-subhalo (right) inferences, respectively. Note that the disagreement is significantly smaller in the case of a transdimensional approach, which correctly predicts the mean subhalo convergence field over most of the lens plane and has smaller disagreement close to the subhalos. However, note that even the transdimensional approach cannot fully constrain the true subhalo convergence field in the vicinity of most subhalos. This is expected because the simulated photon count map contains very little information on their properties.

\begin{figure*}
    \includegraphics[width=\textwidth, trim=0 0 0 0cm, clip]{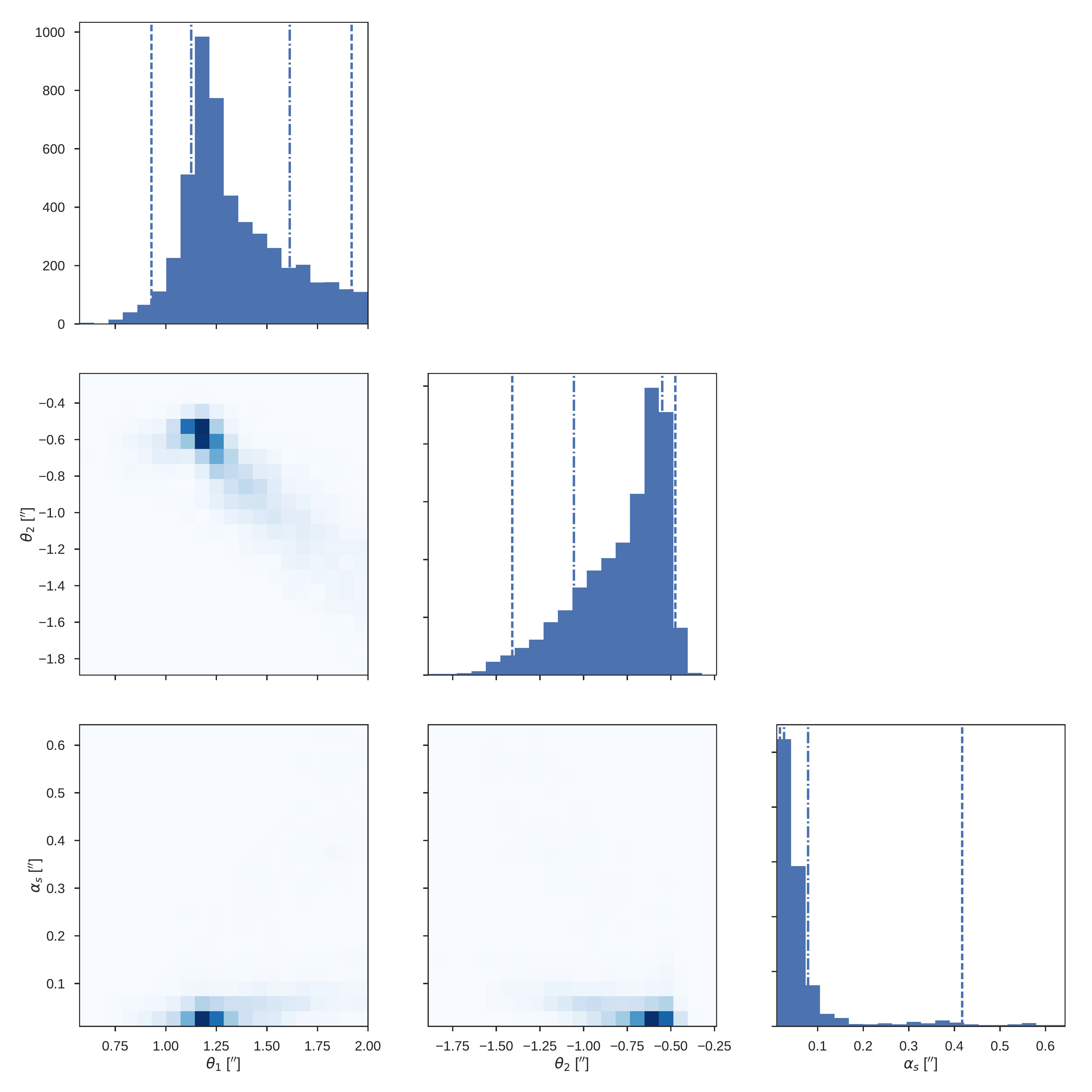}
    \caption{Joint posterior distribution of the horizontal and vertical positions and the deflection strength of the subhalo in the one-subhalo inference. Dashed lines have the same meaning as in Figure \ref{figr:lgalhost_hist}. For comparison, the position of the nearest true subhalo is (1.2, -0.5) arcsec and it has a deflection strength of 0.05 arcsec.}
    \label{figr:listelemfrst_gridoneh}
\end{figure*}

We further show the posterior distribution of the positions, and deflection strength of the model subhalo in the one-subhalo inference in Figure \ref{figr:listelemfrst_gridoneh}. It shows that the joint distribution is highly skewed with heavy tails. Furthermore, the posterior median of the corresponding inferred mass of the subhalo, $4.5\times10^8 M_\odot$, is biased low compared to the mass of the closest true subhalo at (1.2, -0.5) arcsec, $8\times10^8 M_\odot$.

\begin{figure*}
    \subfigure[Nominal]     {\includegraphics[width=.49\textwidth, trim=0 0 0 0cm, clip]{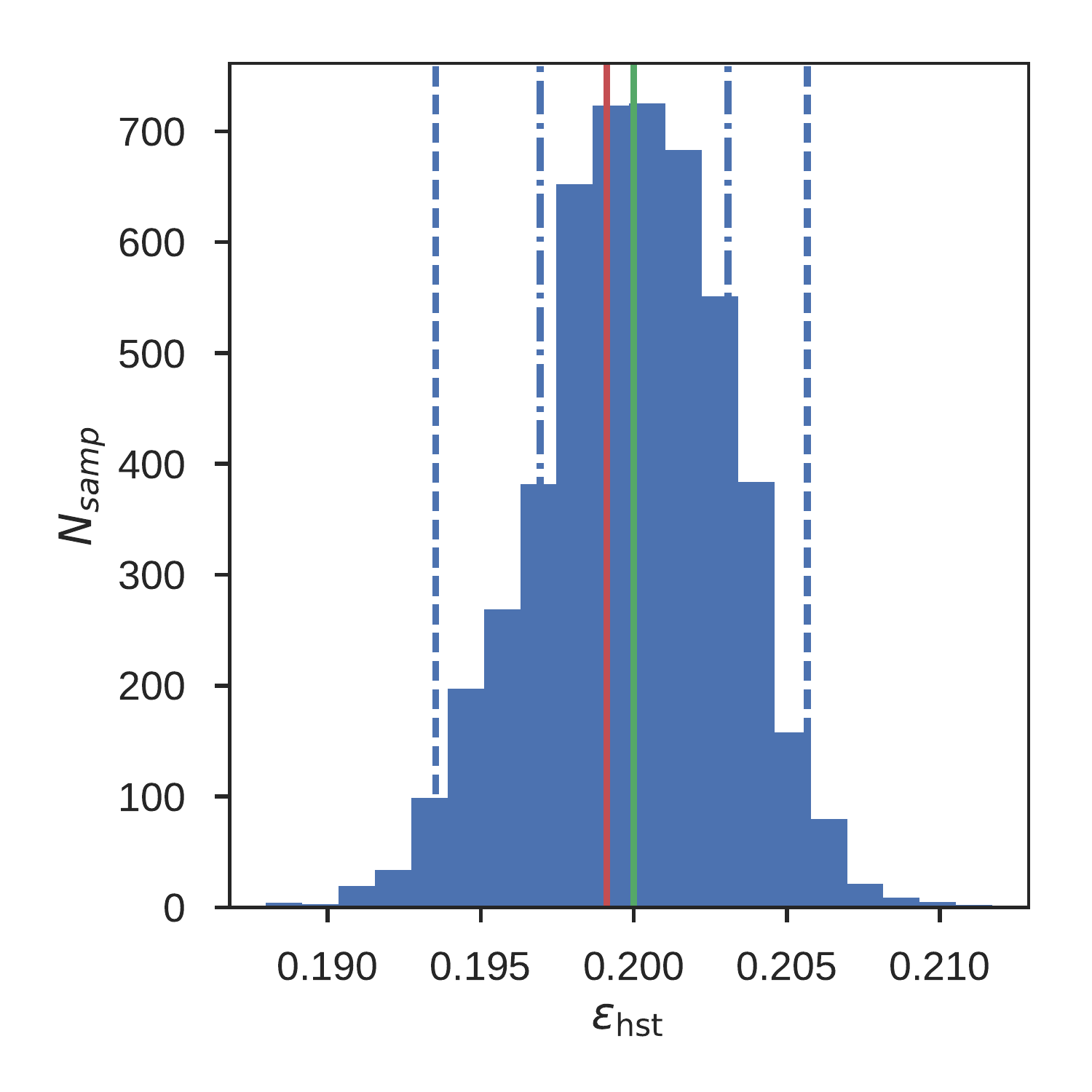}}
    \subfigure[One-subhalo] {\includegraphics[width=.49\textwidth, trim=0 0 0 0cm, clip]{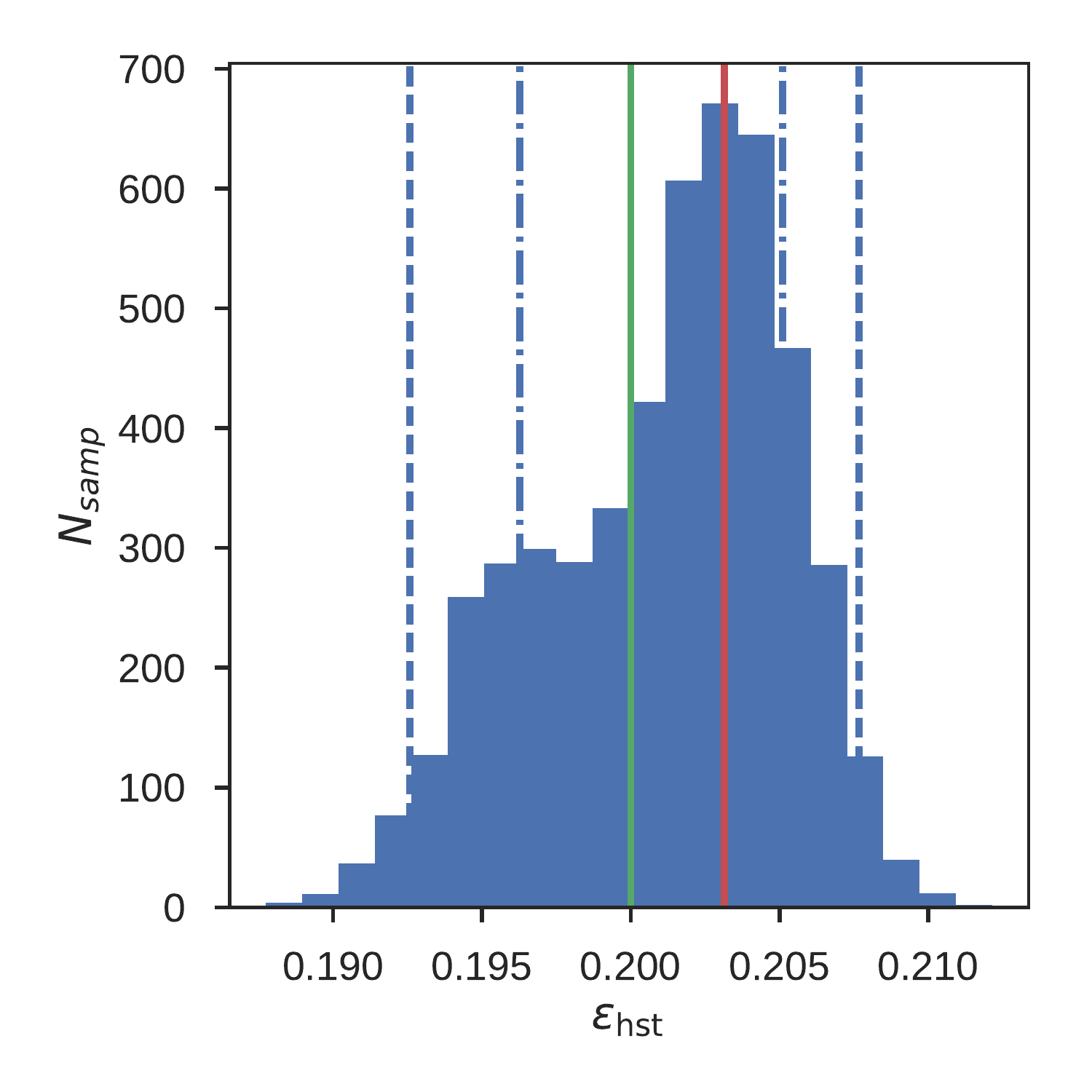}}
    \caption{Histogram of samples from the posterior distribution of the host halo ellipticity. The legend is same as that in Figure \ref{figr:lgalhost_hist}.}
    \label{figr:ellphost_hist}
\end{figure*}

Another problem with fitting the lens image using a fixed number of subhalos is that the posterior distribution on the macro lens model can be significantly biased, depending on the orientation of the non-modeled subhalos. This is because the macro lens model compensates for the absence of subhalos below the detection threshold, resulting in a seemingly good fit that mismodels the underlying true lens metamodel. An example is shown in Figure \ref{figr:ellphost_hist}, where the posterior distribution of the host halo ellipticity is plotted for the cases of nominal (transdimensional) and one-subhalo inferences, using a similar but independently drawn data set. One-subhalo inference is observed to generate multimodality in the posterior distribution of the ellipticity of the host.

\subsection{Extending the metamodel to lower subhalo masses}
\label{sect:lowm}

The lower deflection strength limit of probabilistic cataloging can be made smaller at the expense of making the posterior prior-dominated (thus resulting in no information gain) and significantly increasing the computation time as required for MCMC convergence. Therefore, given the observed data, there is a reasonable lower limit to the deflection strength, e.g., 0.01 arcsec, that can balance information gain and computational complexity.

\begin{figure*}
    \subfigure[Nominal]                     {\includegraphics[width=.49\textwidth, trim=0 0 0 0cm, clip]{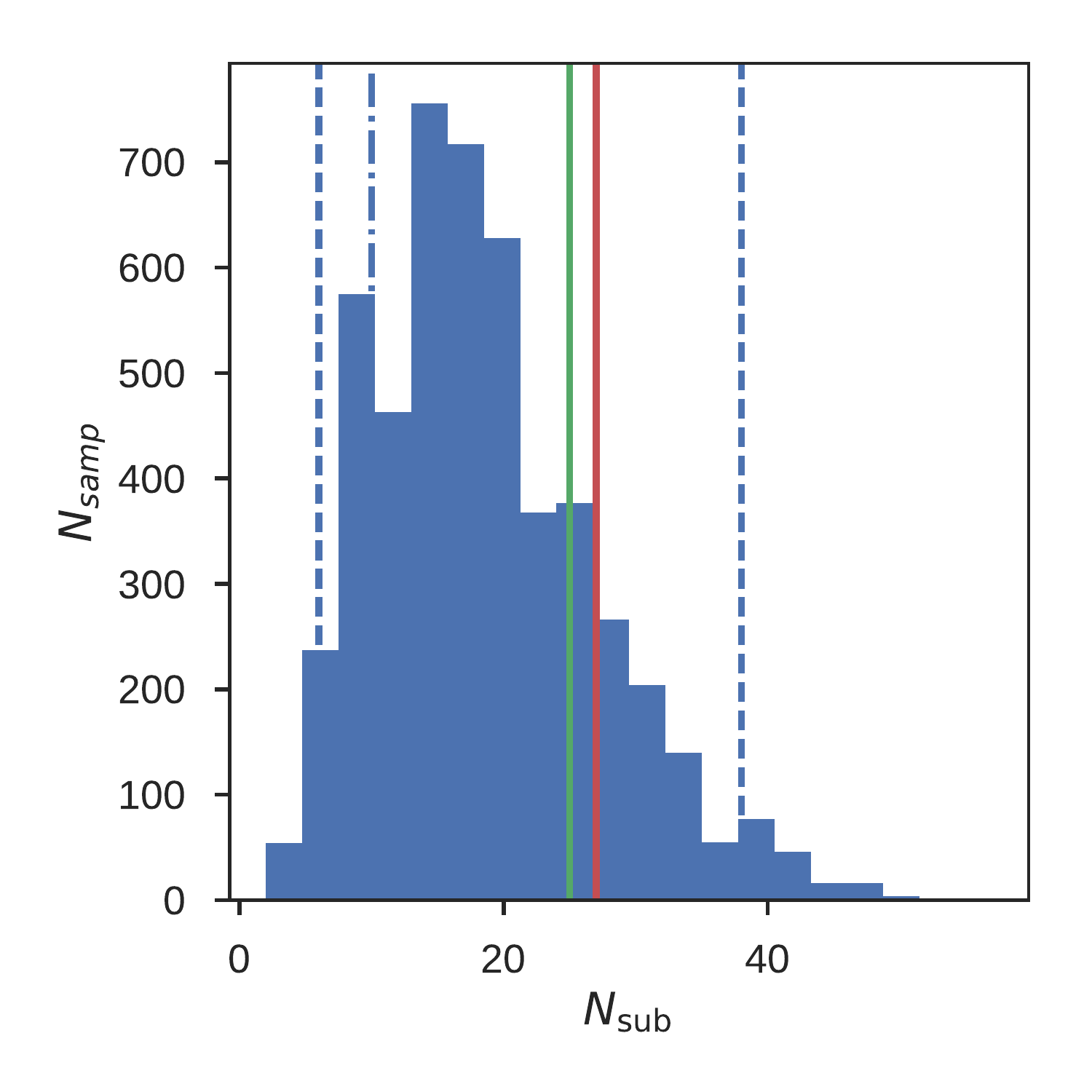}}
    \subfigure[Smaller lower limit on deflection strength] {\includegraphics[width=.49\textwidth, trim=0 0 0 0cm, clip]{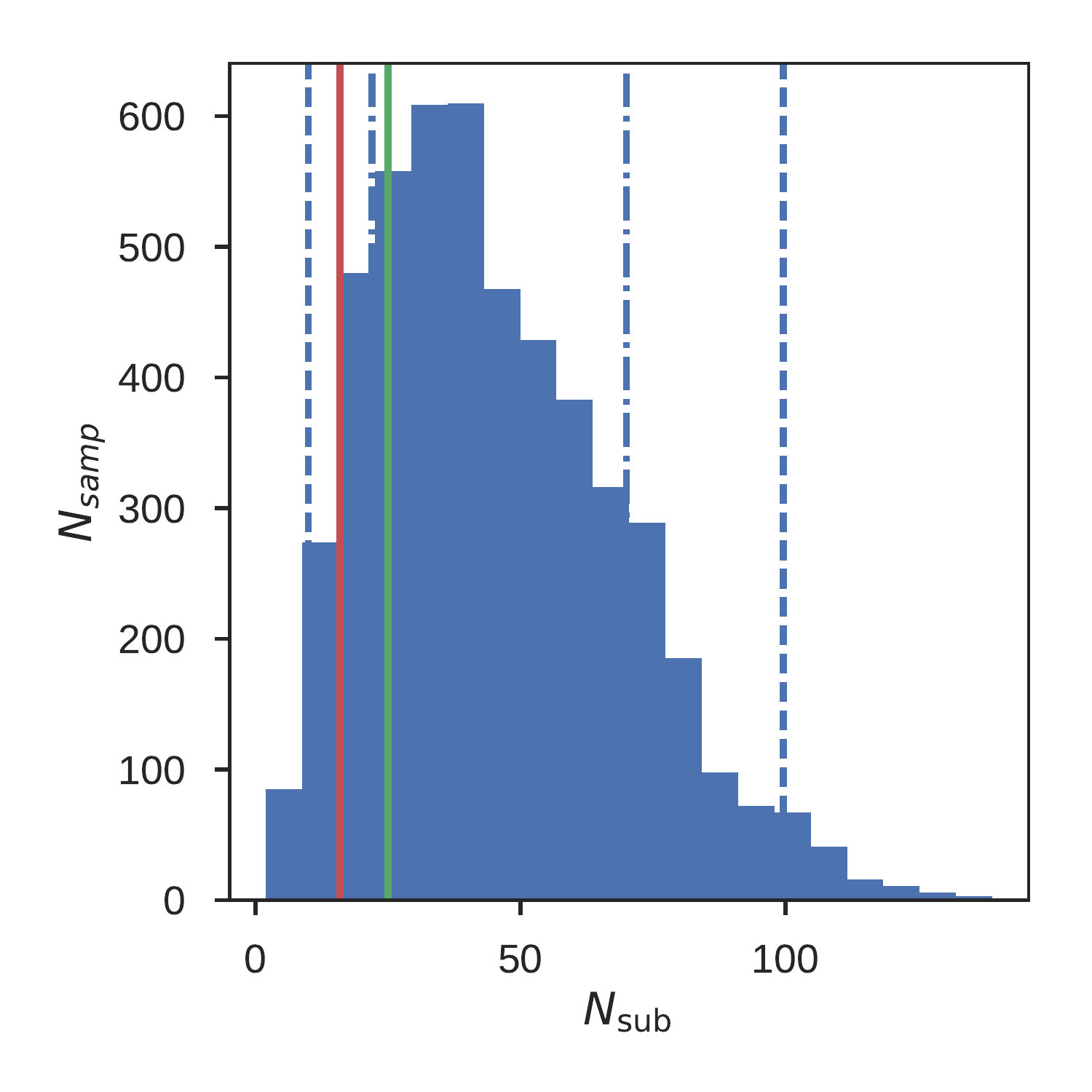}}
    \caption{Histogram of samples from the posterior distribution of the number of subhalos shown with blue. The legend is same as that in Figure \ref{figr:lgalhost_hist}.}
    \label{figr:numbpnts_hist}
\end{figure*}

Nevertheless, for self-consistency, we illustrate an inference where there is no mismodeling at the low end of the subhalo mass function, unlike our nominal results in Section \ref{sect:nomi}, where the minimum subhalo deflection strength allowed in the fitting metamodel was a factor of three higher than the true minimum subhalo deflection strength. Using the same data set as shown in Section \ref{sect:nomi}, Figure \ref{figr:numbpnts_hist} compares the posterior distribution of the number of subhalos in our nominal results (left) and in a run where $\alpha_{\rm{s,min}} = 0.003$ arcsec (right). Given a lower $\alpha_{\rm{s,min}}$, the run shown in the right panel explores a higher number of subhalos whose masses are lower, on average. Note that the two posteriors are consistent with each other. Because we have a uniform prior on the expected number of subhalos, $\mu_{\rm{sub}}$, the fact that there is a preferred scale in the number of subhalos -- and it is consistent with the true number of subhalos -- implies that the posterior is informed by the underlying subhalos below the detection threshold.

\section{No-signal (null) test}
\label{sect:null}

\begin{figure}[b!]
\begin{center}
    \includegraphics[width=0.5\textwidth]{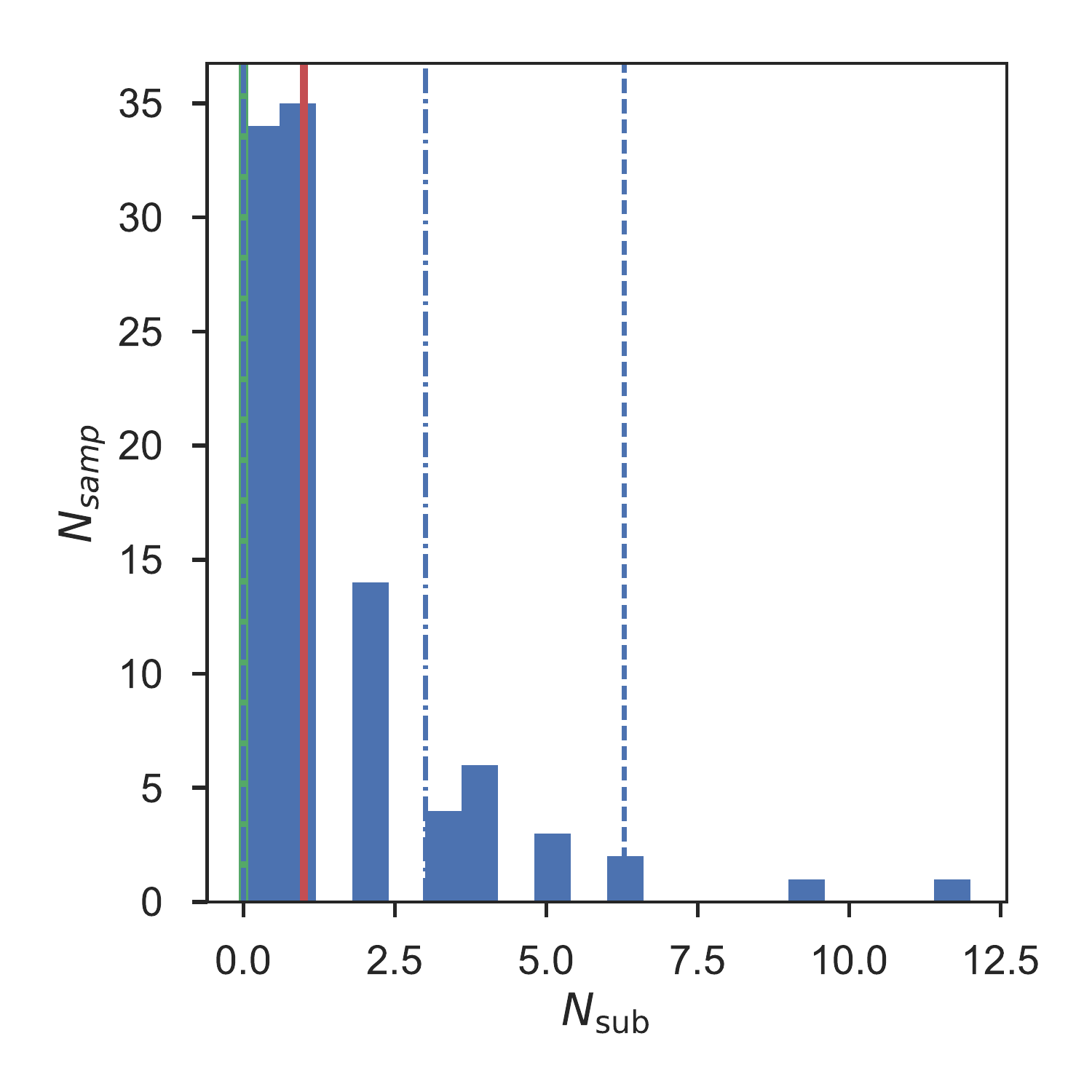}
    \caption{Histogram of the number of subhalos in the posterior when there is no true subhalo that affects the mock image. The legend is same as that in Figure \ref{figr:lgalhost_hist}.}
    \label{figr:numbelemzero}
  \end{center}
\end{figure}

\begin{figure}[b!]
\begin{center}
    \includegraphics[width=0.5\textwidth]{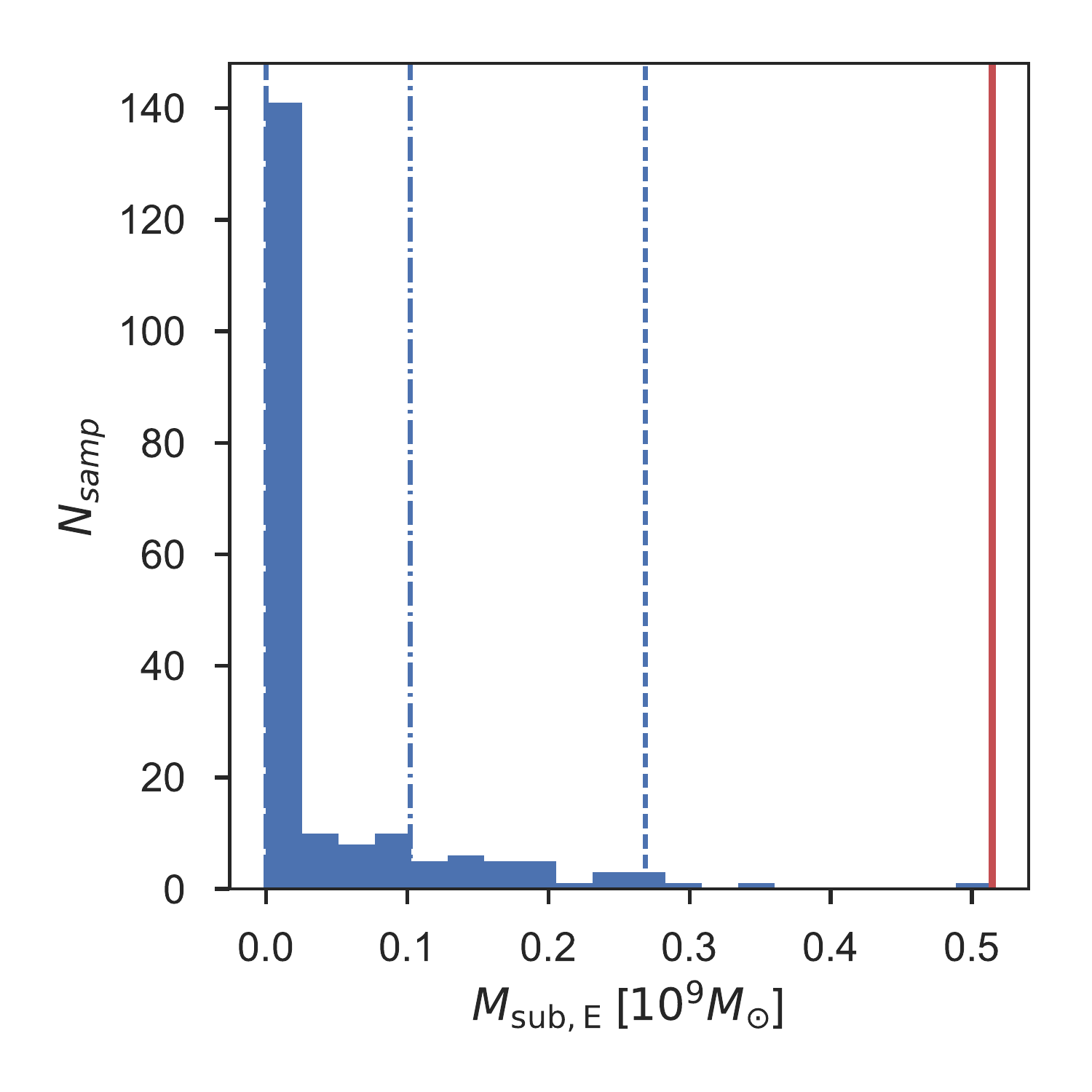}
    \caption{Posterior distribution of the total mass in subhalos inside the Einstein radius of the host galaxy, when there is no true subhalo that affects the mock image. The legend is sames as that in Figure \ref{figr:lgalhost_hist}.}
    \label{figr:histmasssubhtruenone}
  \end{center}
\end{figure}

Any inference framework is expected to produce null results when subject to an input data set that does not contain the signal of interest. We therefore tested our inference framework on a mock data set that was not affected by any subhalo. For this purpose, we generated another mock image as a Poisson draw of an image forward-modeled by the macro lens model, background emission, and PSF (same as that used to generate the main data set presented in the paper), but without any subhalo. We then run \texttt{PCAT} on this image. Figure \ref{figr:numbelemzero} shows the histogram of the number of subhalos in the posterior, which is consistent with the true value of 0. Furthermore, none of the subhalos in the posterior ensemble of catalogs is more significant than 4$\sigma$. The fact that there is a non-zero number of subhalos in the posterior is largely a consequence of the degeneracy between the mass of the main deflector and the total mass of subhalos inside the critical curve. The total mass in the subhalos, as shown in Figure \ref{figr:histmasssubhtruenone}, is also mostly zero, although the posterior does contain some mismodeled subhalo mass due to degeneracies in the lensing problem. The 84th percentile (vertical blue dashed-dotted line) extends to about 3$\times10^8 M_\odot$, which coincides with the maximum likelihood sample (vertical red line), indicating that these few samples overfit the image. 

\section{Sensitivity to the choice of $\alpha_{\rm{s},min}$}
\label{sect:minmdefs}

An important nuisance parameter of our model is the minimum deflection strength allowed by the prior. This is critical because the subhalo mass function is expected to be steep, causing the posterior total number of subhalos to sensitively depend on the minimum deflection strength allowed by the metamodel. The lower $\alpha_{\rm{s},min}$ is allowed to go, the more prior-dominated the posterior becomes. Conversely, taking $\alpha_{\rm{s},min}$ too high causes mismodeling because those subhalos with marginal significances that could fit subtle features in the data are thus removed from the metamodel.

\begin{figure*}
    \includegraphics[width=0.45\textwidth]{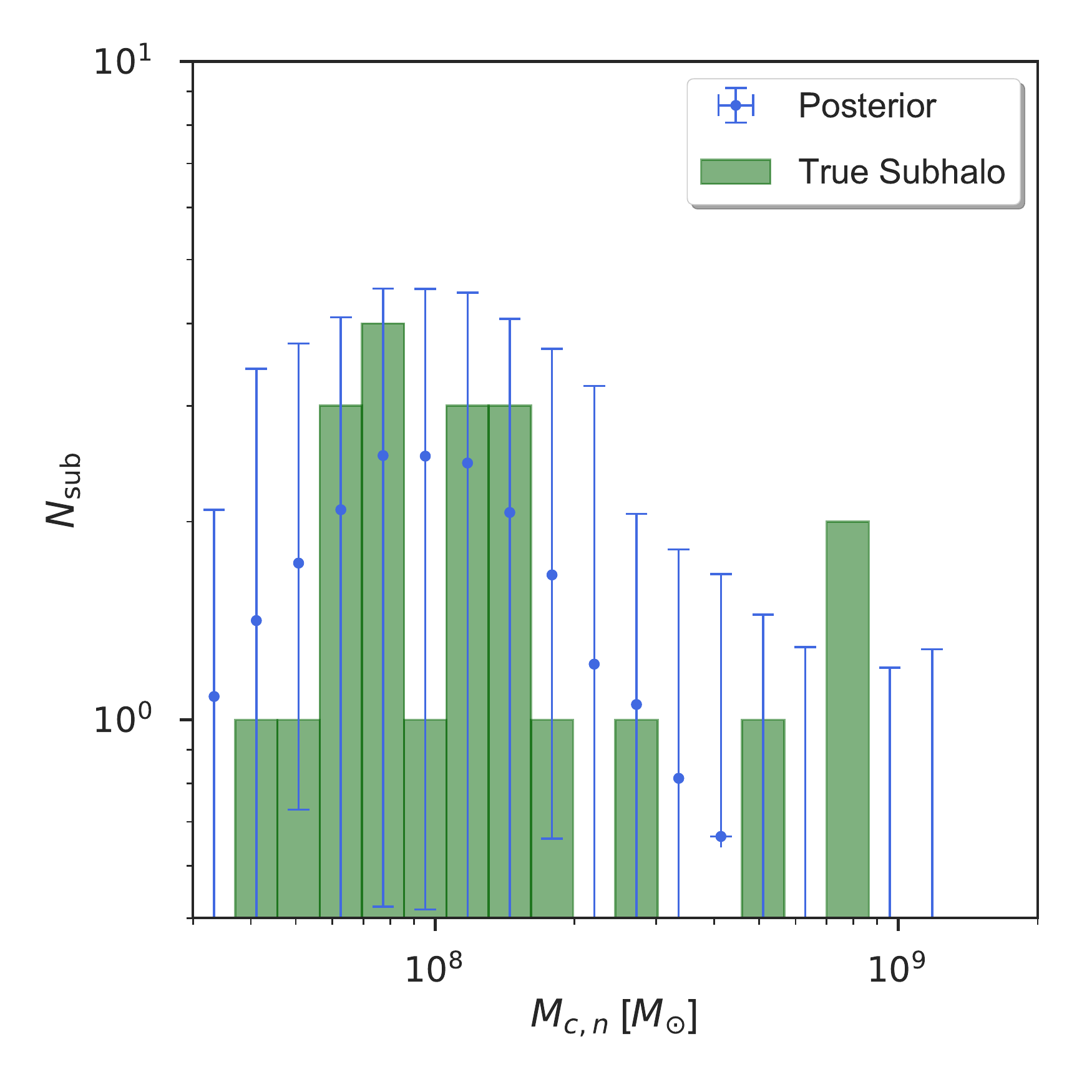}
    \includegraphics[width=0.45\textwidth]{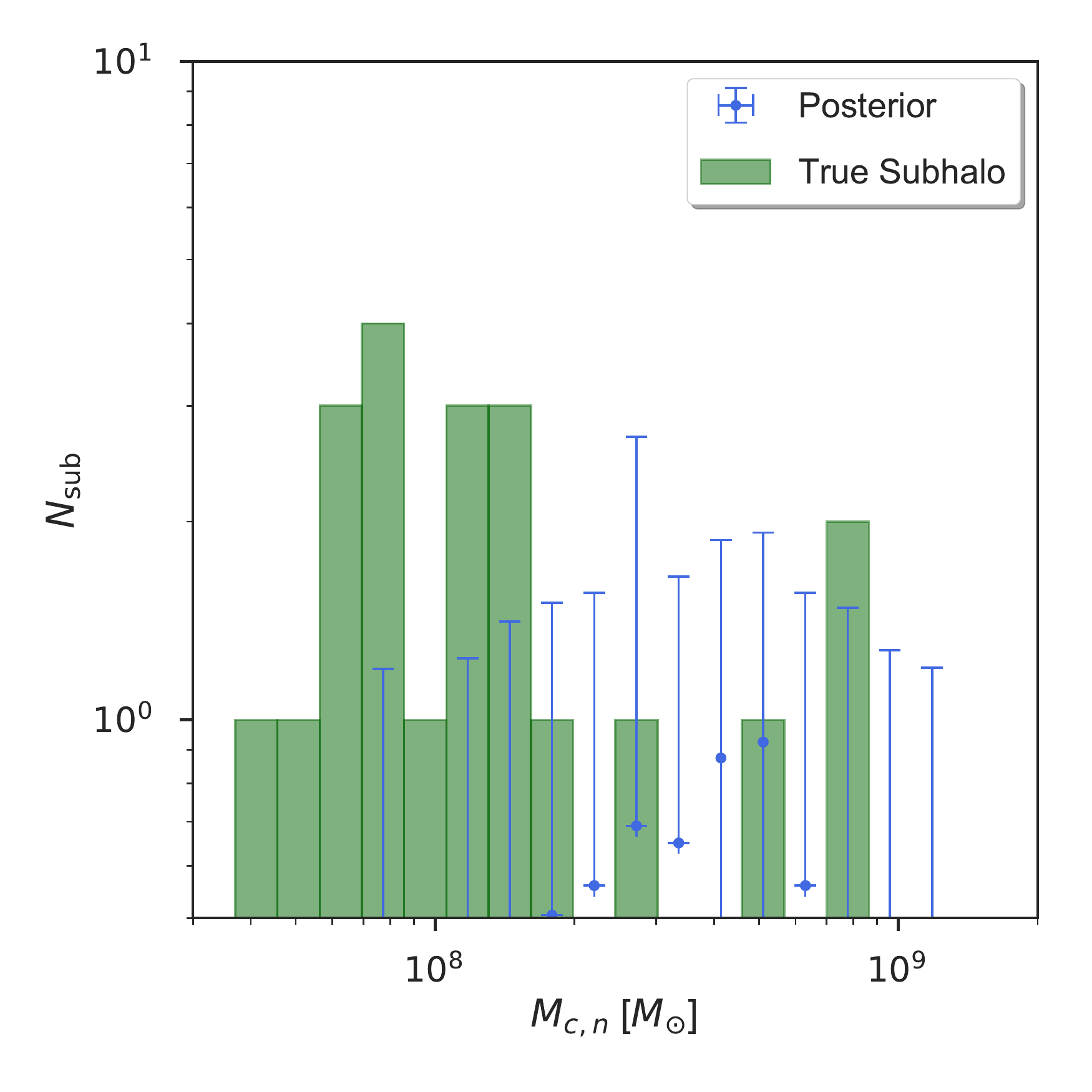}
    \caption{Posterior histogram (blue) of the subhalo masses, when $\alpha_{\rm{s},min}$ is 0.005 arcsec (left) and 0.02 arcsec (right). The green histograms show the mass function of the true subhalos.}
    \label{figr:pdfnhistmcutpop0reg0vari}
\end{figure*}

In Figure \ref{figr:pdfnhistmcutpop0reg0vari} we illustrate the effect of varying this nuisance parameter away from the nominal value of 0.01 arcsec. In the left panel, $\alpha_{\rm{s},min}$ is reduced to 0.005 arcsec, whereas it is increased to 0.02 arcsec in the right panel. Although the posterior in the left panel agrees better with the ground truth at the low-mass end, this happens at the expense of oversplitting the most massive subhalos. Furthermore, the lower end of the posterior subhalo mass function becomes prior-dominated, and could miss a potential cutoff or flattening required by the data. The 0.02 arcsec run on the right performs comparably at the high masses, but fails to explore the subhalo mass function at the low-mass end. 

\section{Discussion}
\label{sect:disc}

Our results using mock data show that probabilistic cataloging is an effective tool to probe lens models with a variable number of deflectors. Our approach significantly differs from conventional cataloging methods because we do not discard information or bias the posterior by imposing a detection threshold on model subhalos, i.e., requiring that model subhalos improve the goodness of fit above some threshold. By subsequent marginalization over the catalog space of subhalos, we obtain robust estimates of the underlying mass model and population characteristics of subhalos, without necessarily individually detecting them. This contrasts with the conventional cataloging, where a few formally detected subhalos and nondetections are used to constrain the subhalo mass function. However, because these detections are not fair draws from the underlying subhalo mass function, this approach cannot account for covariances in the subhalo catalog space, potentially yielding biased results with erroneously small uncertainties.

Another key ingredient of probabilistic cataloging is the exploration of across-model covariances, which is not accounted for in conventional cataloging. For example, given an observed lensed image with a single underlying subhalo, the resulting photon count features can be fitted by a single high-mass subhalo, or multiple lower-mass subhalos. Moreover, because of the inherent degeneracies of the lensing problem, an uncountable set of other configurations can be consistent with the observed data, depending on the positions of subhalos with respect to the lensed emission. Point estimates in the hypothesis space can therefore lead to biased results. Probabilistic cataloging offers a principled way of characterizing these covariances and propagating uncertainties.

In principle, probabilistic cataloging of subhalos could be used to sample arbitrarily low-mass subhalos. However, because the number of subhalos to be modeled steeply increases toward lower subhalo masses, convergence of chains with many low-mass subhalos becomes an issue. Furthermore, because current data sets can only allow the detection of only the few most massive subhalos, if any, the constraints on the subhalo mass function at the low-mass end are largely driven by the prior because the likelihood is essentially flat there.

Currently, the most extensive and homogeneous collection of optical images of galaxy-galaxy type strongly lensed systems is the SLACS collection \citep{Bolton2005, Bolton2008}. These targets have been spectroscopically selected from the Sloan Digital Sky Survey (SDSS) data set conditional on encountering multiple strong emission lines at a redshift higher than that of the target galaxy. They were later followed up by the ACS detector on the HST. However, the limited depth of the HST only allows the most massive subhalos, if any, to be constrained \citep{Vegetti2010, Vegetti2012}. Next-generation, wide-field, high angular resolution telescopes such as WFIRST and EUCLID are expected to revolutionize studies of dark matter substructure. The number of high-SNR strong lens images is expected to go up to thousands \citep{Treu2010,Oguri:2010ns}. Furthermore, greater exposure depth should allow constraints on multiple subhalos per strong lens, making it possible to constrain the subhalo mass function to lower masses. These infrared (IR) telescopes will also carry spectrometers that, in synergy with the photometric redshifts provided by LSST, will allow simultaneous constrains on the redshifts of the model sources and deflectors. Follow-up of high-SNR systems with the JWST and interferometry at submillimeter wavelengths, e.g., ALMA, may allow constraints on the subhalo properties of systems at high redshift.

The current framework we present in this paper can be improved in several ways. First, multi-band images can be incorporated into the framework in a straightforward manner, by introducing color parameters for the source and host emissions. This would make available information contained in the different colors of sky background, source, and foreground galaxies. Furthermore, the source plane emission can be parameterized on an adaptively refined grid, instead of imposing that the source plane emission is from a galaxy with a perfect S\'ersic profile. This would yield a more realistic model of the source plane emission, and a more principled method to marginalize over uncertainties in the emission from the source plane. This is especially relevant for late-type background galaxies, where emission has strong spatial features, such as outflows, bright spots, or dust obscured regions on the disc. We leave this study, along with the application of our method to real data, to future work.

Probabilistic cataloging can also be extended to use temporal information by sampling from the posterior distribution of a lens metamodel given exposures of the strongly lensed system taken at \emph{different} times. Because lensing is also sensitive to the angular diameter distances to the lens and source planes, this would allow an independent measurement of the Hubble constant, $H_0$, as in \cite{Suyu2016}, but with constraints that have been marginalized over dark subhalos that can potentially bias the macro lens parameters. Also, combining multiple exposures may improve sensitivity to substructure via time-delay lensing \citep{Keeton:2009ab,Cyr-Racine2016}, although microlensing could potentially bias this inference \citep{Tie2017}.

Furthermore, probabilistic cataloging provides a principled way to gather information from \emph{independent} strong lens systems, when constraining the probability distributions of subhalo properties. Given images of different strong lenses, the catalog space consistent with these images can be jointly sampled with a \emph{common} hyperparameter set characterizing the population characteristics of the subhalos. This is a significant improvement with respect to traditional cataloging, where incorporating nondetections into the inference may become ambiguous because fixed-dimensional approaches do not consistently account for the fact that non-detections can contain real subhalos or that detections may contain none, two or, more.

In conclusion, we suggest a transdimensional, hierarchical, and Bayesian framework to infer the subhalo mass model and population characteristics in strong lens systems. Using gravitational lensing as a probe, the framework models the lens plane mass distribution and background light emission to draw fair samples from the posterior probability distribution over the lens metamodel, given some photometric data set. The resulting posterior can be marginalized to constrain the subhalo mass fraction and the mass distribution on the lens plane.

\acknowledgments

We thank Christopher W. Stubbs, Daniel Eisenstein, Ana Bonaca, Priyamvada Natarajan, Manoj Kaplinghat, and Stephen K.N. Portillo for useful discussions during the project. We also appreciate useful feedback from Simon Birrer, Simona Vegetti, and Brendon J. Brewer during the review process. This work was performed in part at the Aspen Center for Physics, which is supported by National Science Foundation grant PHY-1607611. F.-Y. C.-R.~acknowledges the support of the National Aeronautical and Space Administration ATP grant NNX16AI12G at Harvard University.

\appendix

\section{Deflection profile of subhalos}
\label{sect:deflsubh}

When projected onto the lens plane, the deflection field due to an untruncated, spherically symmetric NFW subhalo becomes azimuthally symmetric and can be described by a radial profile \citep{Golse2001}
\begin{align}
\begin{split}
\alpha_n &\equiv \theta \frac{\bar{\Sigma}_n(\theta)}{\Sigma_{\rm{crit}}} = \frac{M_{0,n}}{\pi D_{\rm{L}}^2 \Sigma_{\rm{crit}} \theta_{{\rm s},n}} \frac{1}{\theta^\prime} \Big(\ln \frac{\theta^\prime}{2} + F(\theta^\prime)\Big) \\
&= \frac{\alpha_{{\rm s},n}}{\theta^\prime} \Big(\ln \frac{\theta^\prime}{2} + F(\theta^\prime)\Big),
\end{split}
\label{equa:deflsubh}
\end{align}
where $\bar{\Sigma}_n(\theta)$ is the mean surface mass density of the $n$th subhalo inside a subhalo-centric circle of radius $\theta$, and
\begin{align}
\Sigma_{\rm{crit}} = \frac{c^2}{4\pi G} \frac{D_{\rm{S}}}{D_{\rm{LS}}D_{\rm{L}}} 
\end{align}
is the critical surface mass density on the lens plane, which has a value of 3.1$\times 10^9 M_\odot$ kpc$^{-2}$ for our simulated data. For notational brevity, we use the rescaled angular distance, $\theta^\prime \equiv \theta / \theta_{{\rm s},n}$. The function $F(\theta^\prime)$ is given by

\begin{align}
F(\theta^\prime) = 
\begin{dcases}
\dfrac{ \arccos(1/\theta^\prime)}{\sqrt{{\theta^\prime}^2 - 1 }} & \text{if $\theta^\prime > 1$} \\
1 & \text{if $\theta^\prime = 1$} \\
\dfrac{\text{arccosh}(1/\theta^\prime)}{\sqrt{1 - {\theta^\prime}^2}}  & \text{if $\theta^\prime < 1$.}
\end{dcases}
\end{align}
When the truncation is taken into account, the deflection profile becomes \citep{Baltz2009}
\begin{align}
\alpha_{n} &= \frac{\alpha_{{\rm s},n}}{\theta^\prime} \frac{\tau_n^2}{(\tau_n^2 + 1)^2}
\Bigg(\Big(\tau_n^2 + 1 + 2({\theta^\prime}^2 - 1)\Big) F(\theta^\prime) + (\tau_n^2 - 1) \ln \tau_n 
+ \pi \tau_n + \sqrt{\tau_n^2 + {\theta^\prime}^2} \Big(\frac{\tau_n^2 - 1}{\tau_n} L(\theta^\prime) - \pi\Big) \Bigg),
\end{align}
where
\begin{align}
L(\theta^\prime) = \ln \Bigg(\frac{\theta^\prime}{\tau_n + \sqrt{\tau_n^2 + {\theta^\prime}^2}}\Bigg).
\end{align}
In the limit of large $\tau_n = \theta_{{\rm c},n}/\theta_{{\rm s},n}$, this expression reduces to Equation \eqref{equa:deflsubh}. Hence, the three parameters we use to describe the resulting deflection profile of the $n$th subhalo are the normalization $\alpha_{\rm{s},n}$, projected scale, and cutoff radii $\theta_{{\rm s},n}$, $\theta_{{\rm c},n}$.

\section{Markov chain convergence}

\begin{figure}[b!]
\begin{center}
    \includegraphics[width=0.49\textwidth, trim=0 2cm 2cm 0, clip]{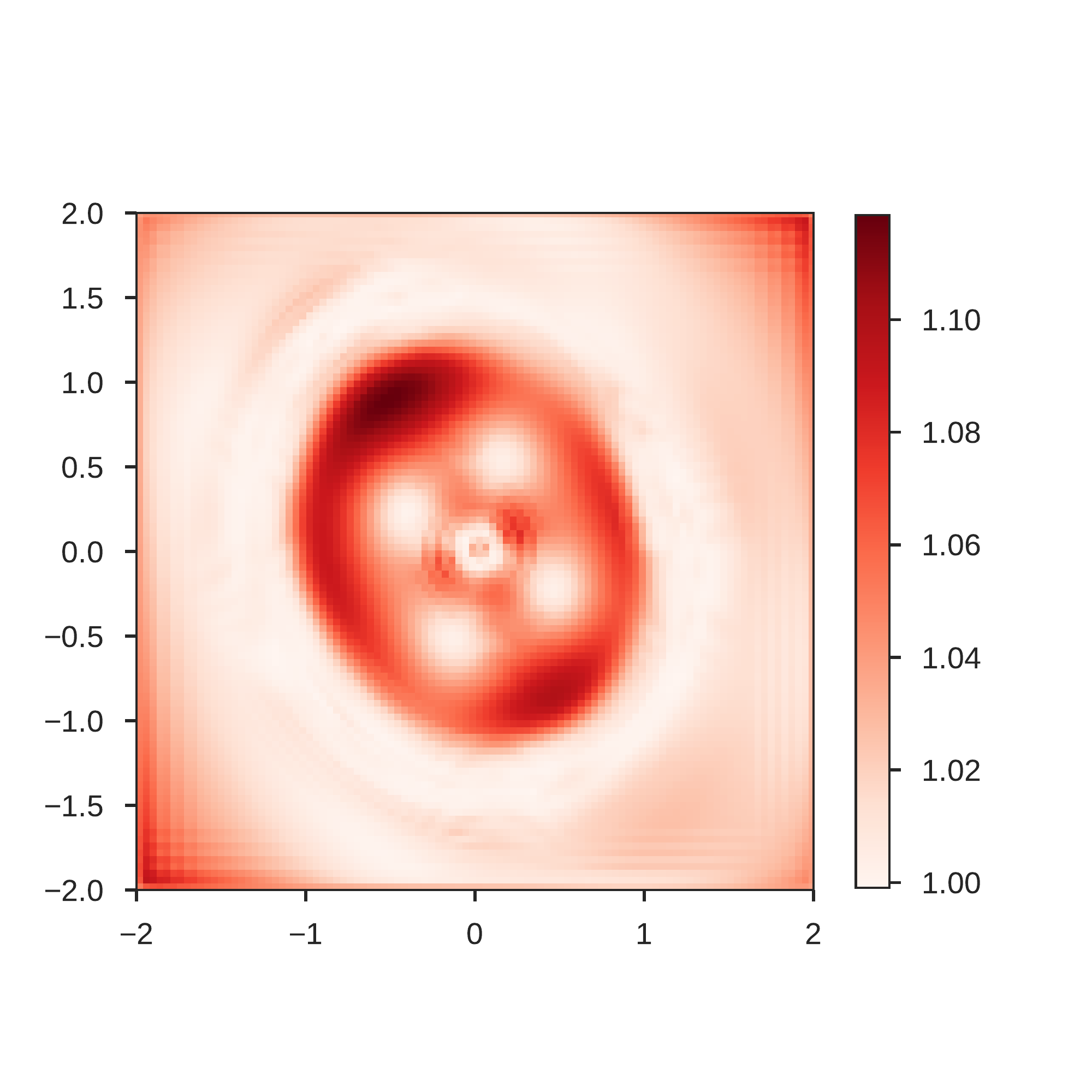}
    \caption{Gelman-Rubin test statistic, i.e., potential scale-reduction factor, evaluated over the posterior samples of the predicted model counts in each pixel.}
    \label{figr:gmrbmaps}
  \end{center}
\end{figure}

When sampling from a target probability distribution by constructing Markov chains, one needs to ensure that all walkers reach the same stationary distribution. Toward this purpose, \texttt{PCAT} monitors the Gelman-Rubin (GR) test statistic, which estimates the factor by which the overall statistical uncertainty of a random variable could be reduced by running the chains longer. Because the chain is transdimensional, however, the statistic is calculated over the data space, i.e., by evaluating the statistic for the number of photons predicted by the metamodel in each pixel.

Figure \ref{figr:gmrbmaps} illustrates the GR test statistic evaluated over the image pixels for a typical \texttt{PCAT} output. At fixed number of samples, the across-chain variance generally increases with the typical number of subhalos. Therefore, given a choice of $\alpha_{\rm{s},min}$, and hence a typical number of subhalos, we ensure that the chain is run long enough to reduce the GR test statistic down to $\lesssim 1.1$. For a typical run with $20-40$ model subhalos, this requires $20 \times 10^6$ samples to be taken before burn-in and thinning. This also implies that the reported constraints have a $\lesssim 10\%$ uncertainty that adds in quadrature to the statistical uncertainty that we derive by assuming that all chains have sampled from the same target distribution, i.e., the posterior distribution over the desired metamodel.

\bibliography{refrdown} 

\end{document}